\documentclass{aastex61}
\usepackage{lineno}
\usepackage{xspace}
\usepackage{soul}
\usepackage{color}
\usepackage{amsmath}

\usepackage{longtable}
\usepackage{threeparttable}

\newcommand{\Fermi}{\emph{Fermi}\xspace}

\newcommand{\Swift}{\emph{Swift}\xspace}

\shorttitle{ \Fermi and \Swift Observations of GRB~190114C}

\begin{document}
\AuthorCallLimit=999
\title{\Fermi and \Swift Observations of GRB~190114C: Tracing the Evolution of High-Energy Emission from Prompt to Afterglow}

\correspondingauthor{Daniel Kocevski, Donggeun Tak, P\'{e}ter Veres, Makoto Arimoto, Ramandeep Gill, Nicola Omodei, }
\author{M.~Ajello}
\affiliation{Department of Physics and Astronomy, Clemson University, Kinard Lab of Physics, Clemson, SC 29634-0978, USA}
\author{M.~Arimoto}
\email{arimoto@se.kanazawa-u.ac.jp}
\affiliation{Faculty of Mathematics and Physics, Institute of Science and Engineering, Kanazawa University, Kakuma, Kanazawa, Ishikawa 920-1192}
\author{M.~Axelsson}
\affiliation{Department of Physics, Stockholm University, AlbaNova, SE-106 91 Stockholm, Sweden}
\affiliation{Department of Physics, KTH Royal Institute of Technology, AlbaNova, SE-106 91 Stockholm, Sweden}
\author{L.~Baldini}
\affiliation{Universit\`a di Pisa and Istituto Nazionale di Fisica Nucleare, Sezione di Pisa I-56127 Pisa, Italy}
\author{G.~Barbiellini}
\affiliation{Istituto Nazionale di Fisica Nucleare, Sezione di Trieste, I-34127 Trieste, Italy}
\affiliation{Dipartimento di Fisica, Universit\`a di Trieste, I-34127 Trieste, Italy}
\author{D.~Bastieri}
\affiliation{Istituto Nazionale di Fisica Nucleare, Sezione di Padova, I-35131 Padova, Italy}
\affiliation{Dipartimento di Fisica e Astronomia ``G. Galilei'', Universit\`a di Padova, I-35131 Padova, Italy}
\author{R.~Bellazzini}
\affiliation{Istituto Nazionale di Fisica Nucleare, Sezione di Pisa, I-56127 Pisa, Italy}
\author{A.~Berretta}
\affiliation{Dipartimento di Fisica, Universit\`a degli Studi di Perugia, I-06123 Perugia, Italy}
\author{E.~Bissaldi}
\affiliation{Dipartimento di Fisica ``M. Merlin" dell'Universit\`a e del Politecnico di Bari, I-70126 Bari, Italy}
\affiliation{Istituto Nazionale di Fisica Nucleare, Sezione di Bari, I-70126 Bari, Italy}
\author{R.~D.~Blandford}
\affiliation{W. W. Hansen Experimental Physics Laboratory, Kavli Institute for Particle Astrophysics and Cosmology, Department of Physics and SLAC National Accelerator Laboratory, Stanford University, Stanford, CA 94305, USA}
% ...


\begin{abstract}
We report on the observations of gamma-ray burst (GRB) 190114C by the  {\it Fermi Gamma-ray Space Telescope} and the {\it Neil Gehrels Swift Observatory}. The prompt gamma-ray emission was detected by the \Fermi Gamma-ray Burst Monitor (GBM), the \Fermi Large Area Telescope (LAT), and the \Swift Burst Alert Telescope (BAT) and the long-lived afterglow emission was subsequently observed by the GBM, LAT, \Swift X-ray Telescope (XRT), and \Swift UV Optical Telescope (UVOT). The early-time observations reveal multiple emission components that evolve independently, with a delayed power-law component that exhibits significant spectral attenuation above 40 MeV in the first few seconds of the burst.  This power-law component transitions to a harder spectrum that is consistent with the afterglow emission observed by the XRT at later times. This afterglow component is clearly identifiable in the GBM and BAT light curves as a slowly fading emission component on which the rest of the prompt emission is superimposed. As a result, we are able to observe the transition from internal shock to external shock dominated emission.  We find that the temporal and spectral evolution of the broadband afterglow emission can be well modeled as synchrotron emission from a forward shock propagating into a wind-like circumstellar environment.  We estimate the initial bulk Lorentz factor using the observed high-energy spectral cutoff. Considering the onset of the afterglow component, we constrain the deceleration radius at which this forward shock begins to radiate in order to estimate the maximum synchrotron energy as a function of time.  We find that even in the LAT energy range, there exist high-energy photons that are in tension with the theoretical maximum energy that can be achieved through synchrotron emission from a shock. These violations of the maximum synchrotron energy are further compounded by the detection of very high energy (VHE) emission above 300 GeV by MAGIC concurrent with our observations.  We conclude that the observations of VHE photons from GRB~190114C necessitates either an additional emission mechanism at very high energies that is hidden in the synchrotron component in the LAT energy range, an acceleration mechanism that imparts energy to the particles at a rate that is faster than the electron synchrotron energy loss rate, or revisions of the fundamental assumptions used in estimating the maximum photon energy attainable through the synchrotron process. 
\end{abstract}

\section{Introduction} \label{sec:introduction} 

Long gamma-ray bursts (GRBs) are thought to represent a specific subset of supernovae in which high-mass progenitors manage to retain a significant amount of angular momentum such that they launch a relativistic jet along their rotation axis at the point of stellar collapse \citep{Woosley1993}. The highly variable emission of gamma rays is thought to be produced by shocks internal to this expanding and collimated outflow \citep{Paczynski1986, Goodman1986, Rees1994}, resulting in the most energetic bursts of electromagnetic emission in the Universe.  This prompt emission is followed by long-lived broadband afterglow emission that is thought to arise from the interaction of the expanding jet with the circumstellar environment \citep{Rees1992, Meszaros1993}. 

Over ten years of joint observations by the {\it Fermi~Gamma-ray Space Telescope} and the {\it Neil Gehrels Swift Observatory} have dramatically expanded our understanding of the broadband properties of both the prompt and afterglow components of GRBs. The \Fermi Gamma-ray Burst Monitor (GBM) has detected over 2300 GRBs in the 11 years since the start of the mission \citep{GBMBurstCatalog_6Years,2FLGC}, with approximately 8$\%$ of these bursts also detected by the \Fermi Large Area Telescope (LAT). These observations have shown a complex relationship between the emission observed by the GBM in the keV to MeV energy range and that observed by the LAT above 100 MeV. The LAT-detected emission is typically, although not always, delayed with respect to the start of the prompt emission observed at lower energies and has been observed to last considerably longer, fading with a characteristic power-law decay for thousands of seconds in some cases \citep{Abdo+09, 2013ApJS..209...11A}; see also the Second LAT GRB catalog \citep[][2FLGC]{2FLGC}. Spectral analysis of the GBM- and LAT-observed emission has shown that it is typically not well fit by a single spectral component, but rather requires an additional power-law component to explain the emergence of the emission above 100 MeV \citep{2009ApJ...706L.138A,Ackermann+11,Ackermann2013,Ackermann2014,2016ApJ...833..139A}.

Simultaneous observations by the X-ray Telescope (XRT) on \Swift of a small subset of LAT detected bursts have revealed that the delayed power-law component observed above 100 MeV is largely consistent with an afterglow origin \cite[e.g.,][]{Ackermann2013}.  This component is commonly observed at X-ray, optical, and radio frequencies, but the extension of the afterglow spectrum to higher energies shows that it is also capable of producing significant emission at MeV and GeV energies. The observation of such a component in the LAT has significantly constrained the onset of the afterglow, allowing for estimates of the time at which the relativistic outflow begins to convert its internal energy into observable radiation. 

In both the prompt and afterglow phases, non-thermal synchrotron emission has long been suggested as the radiation mechanism by which energetic particles accelerated in these outflows radiate their energy to produce the observed gamma-ray emission \citep[see][for reviews]{Piran-99, 2004RvMP...76.1143P}. Evidence for synchrotron emission, typically attributed to shock-accelerated electrons, has been well established through multi-wavelength observations of long-lived afterglow emission \citep{Gehrels2009}. Analysis of GBM observations has also shown that many of the long-standing challenges to attributing the prompt emission to the synchrotron process can be overcome \citep{2011ApJ...741...24B, 2011ApJ...727L..33G, 2013ApJ...769...69B}. Synchrotron emission from shock-accelerated electrons should, in many scenarios, be accompanied by synchrotron self-Compton (SSC) emission, in which some fraction of the accelerated particles transfer their energy to the newly created gamma rays before they escape the emitting region \citep[e.g.,][]{Sari2001, 2008FrPhC...3..306F}. The result is a spectral component that mirrors the primary synchrotron spectrum, but boosted in energy by the typical Lorentz factor of the accelerated electrons.

Despite the predicted ubiquity of an SSC component accompanying synchrotron emission from accelerated charged particles, no unambiguous evidence has been found for its existence in either prompt or afterglow spectra \citep[although see][]{2007ChJAA...7..509W, 2013ApJ...776...95F, 2013ApJ...771L..13T, 2013ApJ...771L..33W}. The LAT detection of only 8\% of 2357 GRBs detected by the GBM (2FLGC) disfavors the ubiquity of bright SSC components in the $0.1-100$ GeV energy range during the prompt emission. When there is detectable emission in the LAT, its delayed emergence, as well as low-energy excesses observed in the GBM data, have likewise disfavored an SSC origin of the prompt high-energy emission above 100 MeV \citep{Abdo2009, Ackermann2011, Ackermann2013}.  Likewise, a recent study by \citet{Ajello2018} has also shown that simultaneous detections of GRB afterglows by \Swift XRT and LAT could be sufficiently well modeled as the high-energy extension of the synchrotron spectrum, with no need for an extra SSC component to explain the late-time LAT-detected emission.  
 
At the same time, there is a maximum energy beyond which synchrotron emission produced by shock-accelerated charged particles becomes inefficient.  This occurs when the shock acceleration timescale approaches the radiative loss timescale, resulting in charged particles that lose their energy faster than they can regain it. This maximum photon energy has been shown to be violated by high-energy photons detected by the LAT from GRB~130427A \citep{Ackermann2014}, including a 95 GeV photon (128 GeV in its rest frame) a few minutes after the burst and a 32 GeV photon (43 GeV in the rest frame) observed after 9 hours. These apparent violations of the maximum synchrotron energy would require an emission component in addition to the shock-accelerated synchrotron emission typically used to model LAT-detected bursts.  SSC  and/or inverse-Compton (IC) emission from the afterglow's forward shock are both expected at TeV energies during the prompt emission, although a spectral hardening and/or a flattening of the LAT light curves is expected as a distinct SSC or IC component passes through the LAT energy range, neither of which was observed in GRB~130427A. In addition, late-time observations by \emph{NuSTAR} provide further support for a single spectral component ranging from keV to GeV energies in GRB~130427A almost a day after the event \citep{2013ApJ...779L...1K}. Synchrotron emission could still be a viable explanation for these observations, but only for an acceleration mechanism that imparts energy to the radiating particles faster than the electron synchrotron energy loss rate, such as through magnetic reconnection.

 Here we report on the high-energy detection of GRB 190114C by the \Fermi GBM and LAT and the \Swift Burst Alert Telescope (BAT), XRT, and UV Optical Telescope (UVOT).  The early-time observations show a delayed high-energy emission above 40 MeV in the first few seconds of the burst, before a transition to a harder spectrum that is consistent with the afterglow emission observed by the XRT and GBM. We find that the temporal and spectral evolution of the broadband afterglow emission can be well modeled as synchrotron emission from a forward shock propagating into a wind-like circumstellar environment.  We estimate the initial bulk Lorentz factor using the observed high-energy spectral cutoff. Considering the onset of the afterglow component, we constrain the deceleration radius in order to estimate the maximum synchrotron energy, which is in tension with high-energy photons observed by the LAT.  The violation of the maximum synchrotron energy is further compounded by the detection of very high energy (VHE) emission above 300 GeV by MAGIC from this burst \citep{2019ATel12390....1M}. We find that the detection of high-energy photons from GRB~190114C requires either an additional emission mechanism at high energies, a particle acceleration mechanism, or revisions to the fundamental assumptions used in estimating the maximum photon energy attainable through the synchrotron process.

The paper is organized as follows. We present an overview of the \Fermi and \Swift instruments in $\S 2$, and a summary of our observations in $\S 3$. The results of our temporal and spectral analyses are described in $\S 4$ and we use those results to model the high-energy afterglow in $\S 5$. We summarize our findings and discuss their implications for future VHE detections in $\S 6$. Throughout the paper we assume a standard $\Lambda$CDM cosmology with $\Omega_{\Lambda}=0.7, \Omega_{M}=0.3, H_0=0.7$. All errors quoted in the paper correspond to 1-$\sigma$ confidence region, unless otherwise noted. 

\section{Overview of Instruments} \label{sec:InstrumentOverview}

\subsection{Fermi GBM and LAT} \label{sec:InstrumentOverview:Fermi}

The {\it Fermi~Gamma-ray Space Telescope} consists of two scientific instruments, the GBM and the LAT. The GBM is comprised of fourteen scintillation detectors designed to study the gamma-ray sky in the $\sim8$ keV to 40 MeV energy range \citep{Meegan2009}. Twelve of the detectors are semi-directional sodium iodide (NaI) detectors, which cover an energy range of 8--1000 keV, and are configured to view the entire sky unocculted by the Earth. The other two detectors are bismuth germanate (BGO) crystals, sensitive in the energy range 200 keV to 40 MeV, and are placed on opposite sides of the spacecraft. Incident gamma rays interact with the NaI and BGO crystals creating scintillation photons, which are collected by attached photomultiplier tubes and converted into electronic signals. The signal amplitudes in the NaI detectors have an approximately cosine response relative to the angle of incidence $\theta$, and relative rates between the various detectors are used to reconstruct source locations.  

 The LAT is a pair-conversion telescope comprising a $4\times4$ array of silicon strip trackers and cesium iodide (CsI) calorimeters covered by a segmented anti-coincidence detector to reject charged-particle background events. The LAT detects gamma rays in the energy range from 20\,MeV to more than 300\,GeV with a field of view (FoV) of $\sim 2.4$ steradians, observing the entire sky every two orbits ($\sim$3 hours) while in normal survey mode.  The deadtime per event of the LAT is nominally 26\,$\mu$s, the shortness of which is crucial for observations of high-intensity transient events such as GRBs.  The LAT triggers on many more background events than celestial gamma rays; therefore onboard background rejection is supplemented on the ground using event class selections that are designed to facilitate the study of a broad range of sources of interest \citep{LATPaper}.
 
\subsection{Swift BAT, XRT, and UVOT} \label{sec:InstrumentOverview:Swift}

The {\it Neil Gehrels Swift Observatory} \citep{Gehrels2005} consists of the BAT \citep{Barthelmy05}, the XRT \citep{Burrows05}, and the UVOT \citep{Roming05}. The BAT is a wide-field, coded mask gamma-ray telescope, covering a FoV of 1.4 sr with partial coding fraction cutoff choice of 50$\%$, and an imaging energy range of 15--150 keV. The instrument's coded mask allows for positional accuracy of 1--4 arcminutes within seconds of the burst trigger. The XRT is a grazing-incidence focusing X-ray telescope covering the energy range 0.3--10 keV and providing a typical localization accuracy of $\sim$1--3 arcseconds.  The UVOT is a telescope covering the wavelength range 170 -- 650 nm with 11 filters and determines the location of a GRB afterglow with sub-arcsecond precision. 

\Swift operates autonomously in response to BAT triggers on new GRBs, automatically slewing to point the XRT and the UVOT at a new source within 1--2 minutes.  Data are promptly downloaded, and localizations are made available from the narrow-field instruments within minutes (if detected).  \Swift then continues to follow-up GRBs as they are viewable within the observing constraints and if the observatory is not in the South Atlantic Anomaly (SAA), for at least several hours after each burst, sometimes continuing for days, weeks, or even months if the burst is bright and of particular interest for follow-up.

\section{Observations} \label{sec:observations} 

On 2019 January 14 at 20:57:02.63 UT ($T_0$), GBM triggered and localized GRB~190114C.  The burst occurred 68$^\circ$ from the LAT boresight and 90$^\circ$ from the Zenith at the time of the GBM trigger.  The burst was especially bright the GBM \citep{2019GCN.23707....1H}, producing over $\sim$30,000 counts per second above background in the most illuminated NaI detector. The LAT detected a gamma-ray counterpart at R$.$A$.$ (J2000), decl.(J2000) = 03$^{\rm h}$38$^{\rm m}$17$^{\rm s}$, $-$26$^\circ$59$^{\prime}$24$^{\prime \prime}$ with an error radius of 3 arcmin \citep{Kocevski2019GCN}. Such a high GBM count rate would normally trigger an Autonomous Repoint Request (ARR), in which the spacecraft slews to keep the burst within the LAT FoV. Unfortunately ARR maneuvers have been disabled since 2018 March 16 due to Sun pointing constraints as a result of an anomaly with one of the two Solar Drive Assemblies that articulate the pointing of the spacecraft's solar panels\footnote{\url{https://fermi.gsfc.nasa.gov/ssc/observations/types/post_anomaly/}}. As a result, the burst left the LAT FoV at $T_0$+ 180 s, and the GBM FoV at $T_0$+260 s when it was occulted by the Earth. The burst re-emerged from Earth occultation at $T_0$ + 2500 s, but remained outside the LAT field of view for an additional orbit, re-entering the LAT FoV at $T_0$ + 8600 s. 

GRB~190114C triggered the \Swift BAT at 20:57:03 UT and the spacecraft immediately slewed to the on-board burst localization \citep{2019GCN.23688....1G}. The XRT began observing the field at 20:58:07.1 UT, 64.63 s after the GBM trigger, with settled observations beginning at $T_{0}$ + 68.27 s. UVOT began observing the field at $T_0$+73.63 s with a 150 s finding chart exposure using a White filter. The XRT and UVOT detected X-ray and optical counterparts, respectively, with a consistent location, with a UVOT position of R$.$A$.$ (J2000), decl.(J2000) = 03$^{\rm h}$38$^{\rm m}$01$^{\rm s}_.$16, $-$26$^\circ$56$^\prime$46$^{\prime \prime}_.$9 with an uncertainty of 0.42 arcsec \citep{2019GCN.23704....1O,2019GCN.23725....1S}, which is also consistent with the LAT position. Both the XRT and the UVOT continued observing the burst location throughout the following two weeks, with the last observation occuring 13.86 days post trigger. The XRT light curve is taken from the XRT GRB light curve repository \citep{2007A&A...469..379E,2009MNRAS.397.1177E}. However, the lower energy limit was raised from the default of 0.3 keV to 0.7 keV in order to avoid an apparent increase in the low-energy background caused by additional events created by the effects of trailing charge on the Windowed Timing (WT) readout mode data (see Section \ref{sec:analysis:spectral:late} and www.swift.ac.uk/analysis/xrt/digest\_cal.php\#trail).

The burst was also detected at high-energies by the MCAL on {\it AGILE} \citep{Ursi2019}, SPI-ACS on {\it INTEGRAL} \citep{Minaev2019}, and Insight-HXMT \citep{Xiao2019}. Most notably the MAGIC Cherenkov telescopes \citep{Mirzoyan2019} also detected the burst, which reported a significant detection of high-energy photons above 300 GeV.  The MAGIC observations mark the first announcement of a significant detection of VHE emission from a GRB by a ground-based Cherenkov telescope.

A host galaxy was identified in Pan-STARRS archival imaging observations by \citet{deUgartePostigo2019} and subsequent spectroscopic observations by \citet{Selsing2019} with the Nordic Optical Telescope found absorption lines in the afterglow spectrum, yielding a redshift of $z = 0.42$. The source was also detected in radio and sub-millimeter \citep{Schulze2019, Tremou2019, Cherukuri2019, Alexander2019, Giroletti2019}. The VLA location of the afterglow as reported by \citet{Alexander2019} was R$.$A$.$ (J2000), decl.(J2000) = 03$^{\rm h}$38$^{\rm m}$01$^{\rm s}_.$191 $\pm$ 0.04 arcsec, $-$26$^\circ$56$^\prime$46$^{\prime \prime}_.$73 $\pm$ 0.02 arcsec, a distance of 4.36 and 0.01 arcmin from the LAT and UVOT locations, respectively.  We adopt this location for the analysis carried out throughout the rest of the paper. 

\section{Analysis} \label{sec:analysis} 

\subsection{Temporal Characteristics} \label{sec:analysis:temporal} 

\begin{figure}[t!]
\begin{center}
\includegraphics[width=0.75\columnwidth]{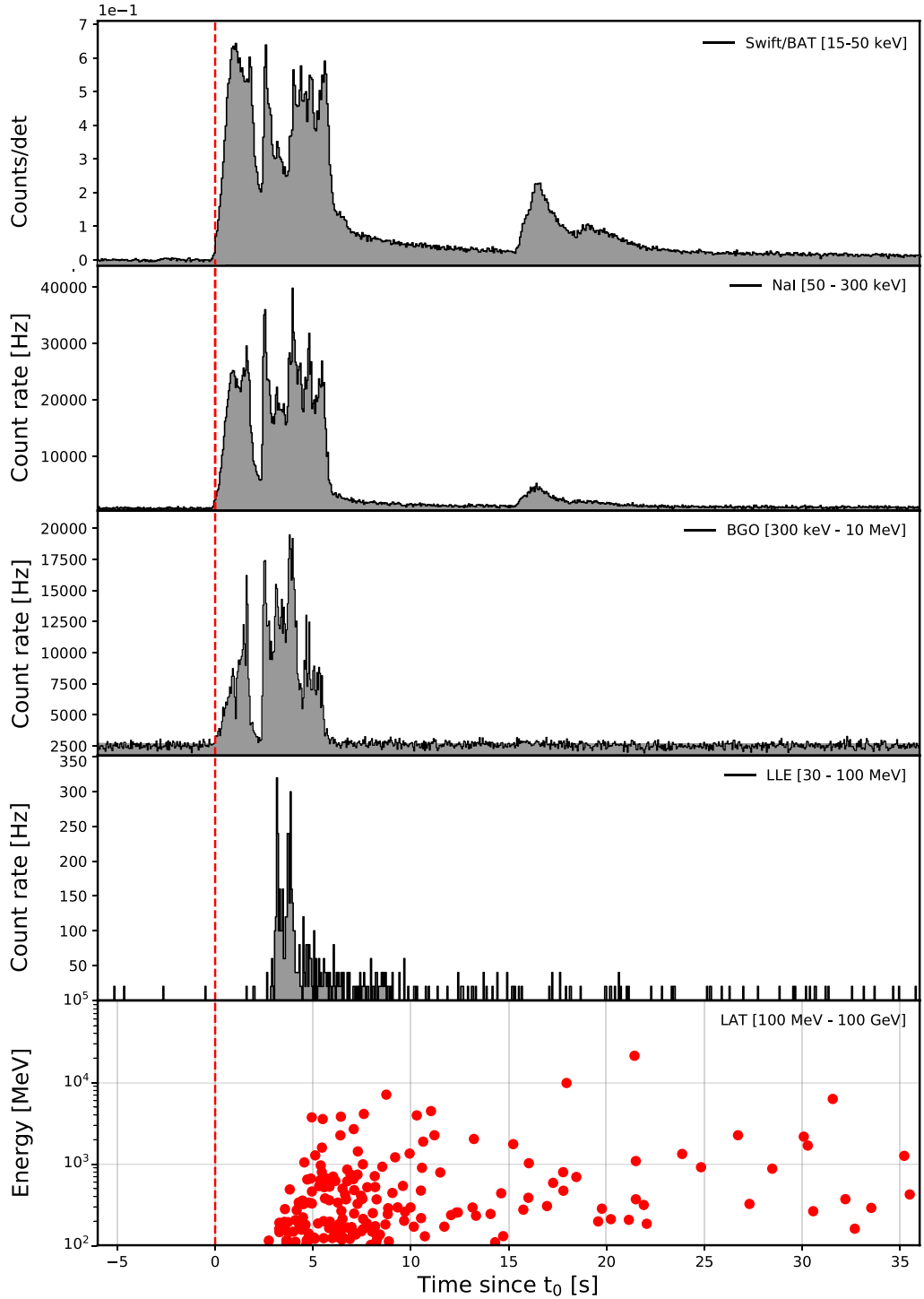}
\caption{Composite light curve for GRB\,190114C: the first panel displays the flux in the 15--50 keV energy range as measured with \Swift/BAT. The second and third panels show the light curves for the most brightly illuminated GBM detectors, NaI (4, 7) and BGO (0) in the 50--300 keV and 0.3--10 MeV energy ranges, respectively. The bottom two panels show the LAT data for the LAT Low Energy (LLE) and  {\tt P8R3Transient010} class events in the 30--100 MeV and $>$100 MeV energy ranges, respectively. In the last panel we show the arrival times and energies of the individual LAT photons with probabilities p$>$0.9 to be associated with the GRB. The red vertical dashed line is the GBM trigger time.}
\label{composite_lightcurve}
\end{center}
\end{figure}

Figure \ref{composite_lightcurve} shows the BAT, GBM, and LAT light curves for GRB 190114C in several different energy ranges.  The BAT and GBM light curves can be characterized by highly variable prompt emission episodes, separated by a quiescent period lasting approximately $\sim$7 s. A strong energy dependence of the light curves is clearly evident, with pulse widths being narrower at higher energies; a feature commonly attributed to hard-to-soft spectral evolution within an emission episode. This trend can be seen to extend up to the LAT Low Energy (LLE) data below 100 MeV \citep{2010arXiv1002.2617P}, although the LAT emission above 100 MeV does not appear to be significantly correlated with the emission at lower energies. Photons with energies $>100$ MeV are first observed at $T_{\rm 0}$ + 2.4 s, consistent with a delayed onset of the high-energy emission seen in other LAT-detected bursts \citep{2FLGC}.  Photons with energies $>1$ GeV are first observed at $T_{\rm 0}$+4.0 s and the highest energy photon was detected at $T_{\rm 0}$+20.9 s with an energy of 21.0 GeV. 

The prompt emission appears superimposed on a smoothly varying emission component that is present during the quiescent period and extends beyond the cessation of the highly variable emission. The $T_{90}$ and $T_{50}$ durations, defined as time intervals within which 90\% and 50\% of the GRB flux was collected, reveal that significant GBM emission above background exists longer than the prompt emission seen within the first 25 s of the burst. We estimate the $T_{90}$ and $T_{50}$ durations, in the 50--300 keV energy range, to be 116.4$\pm$2.6 s and 6.9$\pm$0.3 s, respectively. We also estimate the shortest coherent variation in the light curve, also called the minimum variability time, to be $t_{\rm min}$ = 5.41 $\pm$ 0.13 ms in the NaI detectors, 6.49 $\pm$ 0.38 ms in the BGO detectors and 30.00 $\pm$ 4.74 ms in the LLE band (20--200 MeV) of the LAT detector \citep{2013arXiv1307.7618B}.

\subsection{Spectral Characteristics} \label{sec:analysis:spectral} 
\subsubsection{GBM--LAT Joint Spectral Analysis} \label{sec:analysis:spectral:early}
We examined the underlying spectral characteristics of the prompt emission from GRB 190114C by performing joint time-resolved spectral analysis using the GBM and LAT data from $T_{0}$ to the start of the settled XRT observations at $T_{0}$ + 68.27 s. For GBM, we used the Time-Tagged Event data for two NaI detectors (n4 and n7) from 10 keV -- 1 MeV, and one BGO detector (b0) from 250 keV -- 40 MeV, after considering the spacecraft geometry and viewing angles of the instruments to the burst location. We also include the LLE data, covering an energy range of 30 MeV -- 100 MeV. For both the GBM and LLE data, the background rate for each energy channel was estimated by fitting a second-order polynomial to data before and after GRB 190114C, taking care to exclude a weak soft precursor emission and any extended emission during the power-law decay observed in the GBM.

For the LAT data, we selected {\tt P8R3Transient010} class events in the 100 MeV -- 100 GeV energy range from a region of interest (ROI) of 12$^\circ$ radius centered on the burst location. We applied a maximum zenith angle cut of 105$^\circ$ to prevent contamination from gamma rays from the Earth limb produced through interactions of cosmic rays with the Earth's atmosphere.

We use {\tt gtbin} from the standard ScienceTools (version v11r5p3)\footnote{\url{http://fermi.gsfc.nasa.gov/ssc/}} to generate the counts spectrum of the observed LAT signal and {\tt gtbkg} to extract the associated background by computing the predicted counts from cataloged point sources and diffuse emission components in the ROI.  We draw cataloged point sources from the 3FGL catalog and we use the publicly available\footnote{\url{http://fermi.gsfc.nasa.gov/ssc/data/access/lat/BackgroundModels.html}} isotropic ({\tt gll\_iem\_v06}) and Galactic diffuse ({\tt iso\_P8R2\_TRANSIENT020\_V6\_v06}) templates\footnote{The difference between the P8R2 and P8R3 isotropic spectra are small and do not affect the results of this analysis.} to model the diffuse emission components. The LAT instrument response for the each analysis interval was computed using {\tt gtrspgen}. 

The spectral fits were performed using the \textit{XSPEC} software package (version 12.9.1u) \citep{1996XspecProc}, in which we minimize the $PG_{\rm stat}$ statistic for Poisson data with Gaussian background  \citep[][]{2011hxra.book}. The best-fit model is selected by minimizing the Bayesian information criterion \citep[BIC;][]{schwarz1978}. For each time interval, we test a variety of spectral models, including a power law (PL), a power law with an exponential cutoff (CPL), the Band function \citep[Band;][]{1993ApJ...413..281B}, a black body (BB), and combinations thereof. 

\begin{figure}[t!]
\begin{center}
\includegraphics[width=0.90\columnwidth]{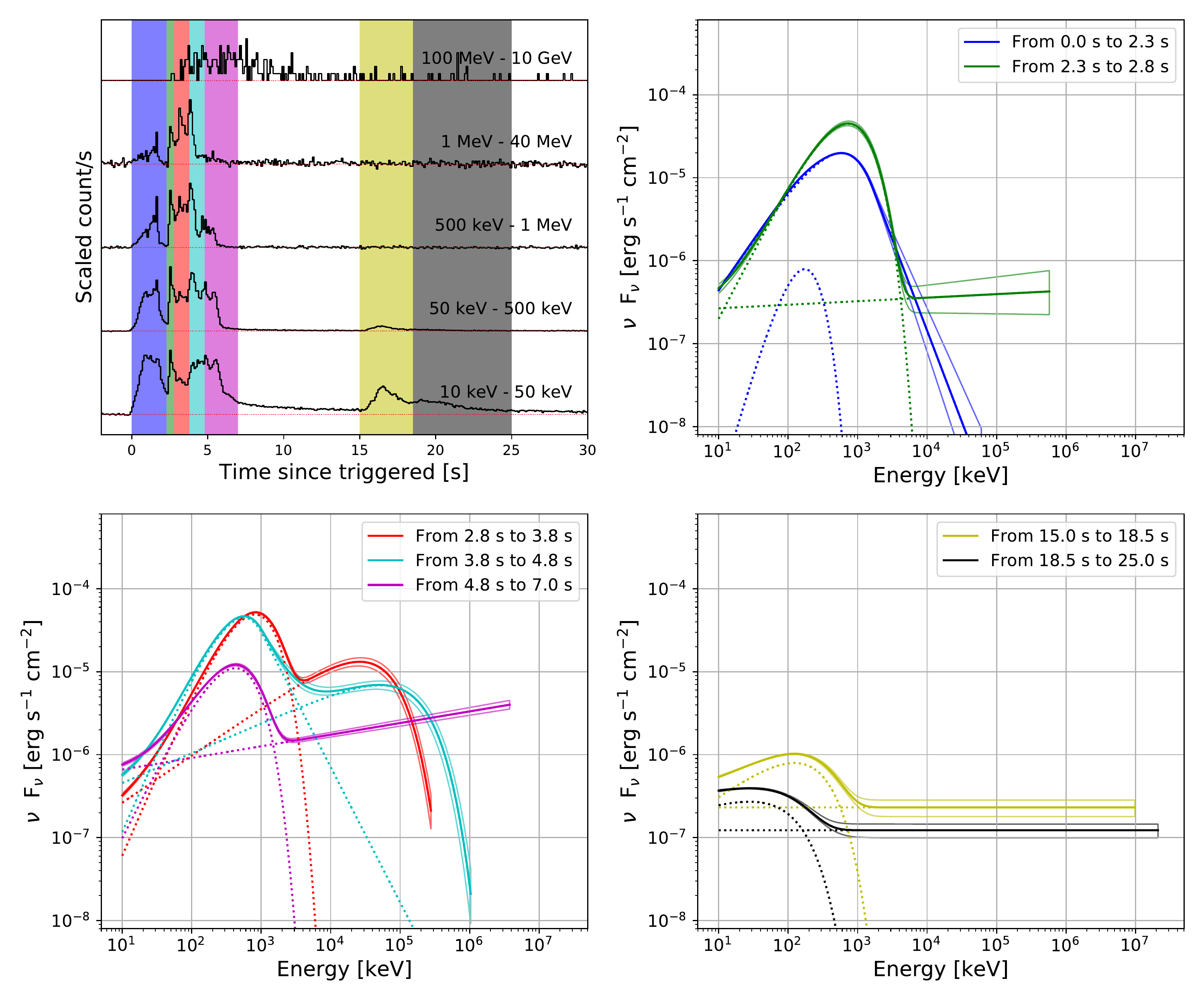}
\caption{The scaled light curves and the $\nu$F$_{\nu}$ model spectra (and $\pm$1$\sigma$ error contours) for each of the time intervals described in Section~\ref{sec:analysis:spectral:early}. Each SED extends up to the energy of the highest-energy photon detected by LAT. The color coding used in the shading of time intervals in the top--left panel is carried over to the energy spectra in the other three panels. The dotted lines represent the components of the model spectra. The best-fit model and its parameters are listed in Table~\ref{tab:Fermi}. }
\label{SED}
\end{center}
\end{figure}

The time interval from $T_{0}$ to  $T_{0}$ + 25 s was subdivided into 7 intervals after considering the temporal characteristics shown in Figure~\ref{SED}. Figure~\ref{SED} also shows the best-fit model for each time interval. The spectrum of the first pulse phase ($T_{0}$ + 0 -- 2.3 s) is best fitted with the Band + BB model. The addition of the BB component to the Band component is weakly preferred ($\Delta$BIC $\sim$ 2). The peak energy ($E_{\rm pk}$) for the Band component is 586 $\pm$ 14 keV, and the temperature of the BB component is 44 $\pm$ 5 keV. The temperature of the BB component is consistent with similar components seen in other bright GRBs \citep{2012ApJ...757L..31A, 2011ApJ...727L..33G, 2013ApJ...770...32G}.

\begin{table*}
\tiny
\centering 
\caption{Spectral fitting to GBM + LLE + LAT data (10 keV--100 GeV) for various time intervals }
	\begin{tabular}{c c c c c c c c c c c c c c}
    \hline\hline
& & & \multicolumn{4}{c}{Main component} & &\multicolumn{4}{c}{Additional component} \\\cline{4-7}\cline{9-12}
From & To & Model\footnote{For the PL, CPL, and Band models, the pivot energy is fixed to 100 keV} & Norm.\footnote{photons cm$^{-2}$ s$^{-1}$ keV$^{-1}$} & $\Gamma_{\rm ph,low}$ & $\Gamma_{\rm ph,high}$ & $E_{pk}$ & & Norm.$^{b}$ &$\Gamma_{\rm ph,PL}$ & $E_{pk}$ & kT & $PG_{\rm stat}/dof$ & BIC\\
{[ s ]} & [ s ] &  & & & & [ keV ] & & & & [ MeV ] &[ keV ]\\\hline\hline
0.0 & 2.3 & Band & 0.518$^{+0.005}_{-0.005}$ & -0.73$^{+0.01}_{-0.01}$ & -4.00$^{+0.27}_{-0.42}$ & 548.6$^{+7.7}_{-7.6}$ & &  &  & & & 518/353&542\\
& & Band+BB & 0.481$^{+0.011}_{-0.011}$& -0.77$^{+0.01}_{-0.01}$ & -4.20$^{+0.31}_{-0.46}$ & 585.4$^{+14.2}_{-13.6}$ & &11.54$^{+5.46}_{-4.30}$  &  & &44.2$^{+4.9}_{-4.7}$ & 505/351&540\\\hline 
2.3 & 2.8 & CPL+PL & 0.555$^{+0.009}_{-0.009}$ & -0.36$^{+0.03}_{-0.03}$ &  & 730.0$^{+16.2}_{-15.5}$ & &0.018$^{+0.004}_{-0.003}$  &-1.96$^{+0.05}_{-0.06}$ & & & 425/352&454 \\\hline 
2.8 & 3.8 &  CPL+PL& 0.374$^{+0.006}_{-0.006}$& -0.09$^{+0.03}_{-0.03}$ &  & 840.8$^{+13.1}_{-12.9}$ & &0.040$^{+0.002}_{-0.002}$ & -1.68$^{+0.01}_{-0.01}$ & & & 769/352&799\\
 & &  CPL+CPL& 0.355$^{+0.007}_{-0.007}$& -0.04$^{+0.03}_{-0.03}$ &  & 814.9$^{+13.4}_{-13.0}$ & &0.061$^{+0.004}_{-0.004}$ & -1.43$^{+0.02}_{-0.02}$ &26.1$^{+2.6}_{-2.3}$ & &477/351& 512
 \\\hline
 3.8 & 4.8 & Band+PL &0.706$^{+0.011}_{-0.011}$ &
 -0.05$^{+0.03}_{-0.03}$ & -3.60$^{+0.19}_{-0.28}$ & 562.8$^{+9.6}_{-9.2}$ & &0.050$^{+0.003}_{-0.003}$ &-1.64$^{+0.02}_{-0.02}$ & & & 577/351&612\\
& & Band+CPL & 0.675$^{+0.010}_{-0.010}$& -0.05$^{+0.03}_{-0.03}$ & -3.63$^{+0.21}_{-0.26}$ & 563.1$^{+8.8}_{-9.6}$ & &0.065$^{+0.004}_{-0.004}$ &-1.64$^{+0.02}_{-0.02}$ & 51.5$^{+9.8}_{-7.4}$& & 519/350&560\\\hline 
4.8 & 7.0 & CPL+PL &0.322$^{+0.006}_{-0.006}$ & -0.30$^{+0.04}_{-0.04}$ &  & 425.4$^{+7.7}_{-7.4}$ & & 0.057$^{+0.002}_{-0.002}$& -1.86$^{+0.01}_{-0.01}$& & & 467/352&494\\\hline
15 & 18.5 & CPL+PL & 0.080$^{+0.005}_{-0.005}$ & -1.41$^{+0.08}_{-0.06}$ &  & 122.9$^{+7.5}_{-6.7}$ & & 0.014$^{+0.003}_{-0.003}$ & -2.00 $_{\rm fixed}$ & & & 407/353&430 \\\hline
 18.5 & 25 & CPL+PL & 0.030$^{+0.005}_{-0.004}$ & -1.74$^{+0.09}_{-0.08}$ &  & 27.7$^{+3.3}_{-4.1}$ & & 0.008$^{+0.001}_{-0.001}$ & -2.00 $_{\rm fixed}$ & & &454/353& 478 \\\hline\hline 
     \end{tabular}
  \begin{tablenotes}
  \item Errors correspond to 1-$\sigma$ confidence region.
  \end{tablenotes}
    \label{tab:Fermi}
\end{table*}

The main spectral component during the brightest emission episode observed from $T_{0}$ + 2.3 s to $T_{0}$ + 7.0 s is characterized by  many short and overlapping pulses and is best fit by either a CPL or Band function. During this phase, the low-energy spectral index is very hard, ranging between $-$0.4 -- 0.0 (see Table~\ref{tab:Fermi}). The peak energy ($E_{\rm pk}$) reaches a maximum value of $E_{\rm pk}$ $\sim$ 815 keV from $T_{0}$ + 2.8 s to $T_{0}$ + 3.8 s, before decreasing in time (see Table~\ref{tab:Fermi}). 

An additional PL or CPL component begins to appear during the $T_{0}$ + 2.3 s to $T_{0}$ + 2.8 s time interval and lasts throughout the prompt emission phase. Arrival of the first LAT events above 100 MeV associated with the source begins at $T_{0}$ + $\sim$2.7 s, consistent with the emergence of this spectral component. In the third ($T_{0}$ + 2.8 s to $T_{0}$ + 3.8 s) and fourth ($T_{0}$ + 3.8 s to $T_{0}$ + 4.8 s) time intervals, this additional component increases in brightness and exhibits a high-energy cutoff which increases in energy with time, ranging from 26 -- 52 MeV (see Table~\ref{tab:Fermi}). The high-energy cutoff is strongly required in both time intervals compared to the models without the high-energy cutoff ($\Delta$BIC $\gg$ 10). After $\sim$ 4.8 s, the high-energy cutoff in this additional component disappears, and the high-energy emission is well described by a PL with a photon index ($dN/dE \propto E^{\Gamma_{\rm ph}}$) of $\Gamma_{\rm ph,PL} = -1.86 \pm 0.01$  or correspondingly an energy index ($F_{\nu} \propto \nu^{\beta}$) of $\beta_{\rm PL} = -0.86 \pm 0.01$.

\begin{figure}[t!]
\begin{center}
\includegraphics[width=0.6\columnwidth]{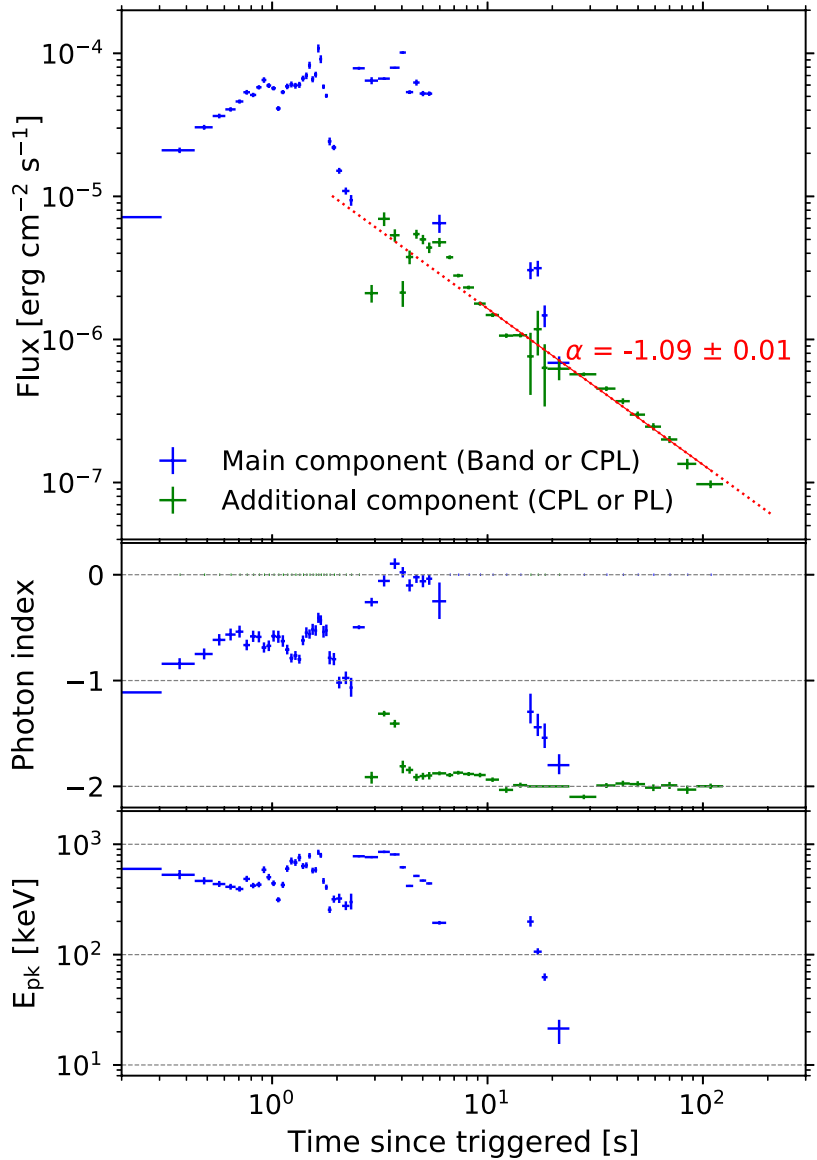}
\caption{Temporal and spectral evolution of each spectral component. \textit{Top Panel}: energy flux in the 10 keV--1 MeV (blue) and 100 MeV--1GeV (green) energy ranges, \textit{Middle Panel}: photon index (for the Band function we refer to the low-energy photon index), and \textit{Bottom Panel}: $E_{\rm pk}$, where we use the trigger time $T_{\rm 0}$}
\label{Component}
\end{center}
\end{figure}

After the bright emission phase, the long-lived extended emission observed by the LAT is best described by a PL with an almost-constant photon index of $\Gamma_{\rm ph,PL} \sim -2$, as shown in Figure~\ref{Component}. Figure~\ref{Component} also shows that the energy flux of this extended emission phase (100 MeV--1 GeV) shows a power-law decay in time  ($F_{\nu} \propto t^{\alpha}$), with an exponent of $\alpha_{\rm LAT} = -1.09 \pm 0.02$. Extrapolation of this extended emission back into the earlier bright emission phase reveals that the flux from the additional spectral component in the prompt emission evolves similarly to the extended emission. This implies that the emission from the additional component and the extended emission may be from the same region. Since the power-law spectral and temporal characteristics of this broadband emission resemble the representative features of GRB afterglows, the end of the bright emission phase at about $\sim$ 7 s represents the transition from the prompt to afterglow-dominated emission.

In addition to the extended emission, a weaker, short-duration pulse, with soft emission primarily below $\lesssim$ 100 keV, is observed from $T_{0}$ + 15 s to $T_{0}$ + 25 s. This weak pulse, along with the long-lasting extended emission, is well described by the CPL + PL model. For these periods, we fix the photon index of the PL component to $-2.0$, assuming that the photon index of the energy spectrum of the extended emission is unchanged in time.

\subsubsection{Fermi--Swift Joint Spectral Analysis \label{sec:analysis:spectral:late}}
We continue the time-resolved spectral analysis from $T_{0}$ + 68.27 s to $T_{0}$ + 627.14 s, but now include \Swift data. For GBM, we prepared the data using the same process as described in Section~\ref{sec:analysis:spectral:early}, although for this time interval we excluded channels below 50 keV because of apparent attenuation due to partial blockage of the source by the spacecraft that is not accounted for in the GBM response. For LAT, we decreased the ROI radius to 10$^\circ$ and increased the maximum zenith angle cut to 110$^\circ$. Both changes are made in order to reduce the loss of exposure that occurs when the ROI crosses the zenith angle cut and begins to overlap the Earth’s limb. This increase in exposure, though, comes at the expense of increased background during intervals when the Earth’s limb is approaching the burst position. The rest of the process is the same as described in Section~\ref{sec:analysis:spectral:early}.

We retrieve \Swift data from the HEASARC archive. The BAT spectra are generated using the event-by-event data collected from $T_{\rm 0,\, BAT}-239$ s to $T_{\rm0,\, BAT}+963$ s, with the standard BAT software (HEASOFT  6.25\footnote{http://heasarc.nasa.gov/lheasoft/}) and the latest calibration database (CALDB\footnote{http://heasarc.gsfc.nasa.gov/docs/heasarc/caldb/swift/}). The burst left the BAT FoV at $\sim T_{\rm0,\,BAT} + 720$ s, and was not re-observed until $\sim T_{\rm0,\,BAT} + 3800$ s. For the intervals that include spacecraft slews, an average response file is generated by summing several short-interval (5 s) response files, weighted by the counts in each interval \citep[see][for a more detailed description]{Lien2016}.

The XRT acquired the source at $T_{0}$ + 64.63 s, and started taking WT data at $T_{0}$ + 68.27 s. In the analysis that follows the XRT data were initially processed by the XRT data analysis software tools available in {\sc HEASOFT} version 6.25, using the gain calibration files released on 2018-Jul-10. Prior to extracting spectra, we processed the WT event data using an updated, but as yet unreleased, version of the XRT science data analysis task {\sc xrtwtcorr} (version 0.2.4), which includes a new algorithm for identifying unwanted events caused by the delayed emission of charge from deep charge traps that have accumulated in the CCD due to radiation damage from the harsh environment of space. Such trailing charge appears as additional low-energy events and can cause a significant spectral distortion at low energies, especially for a relatively absorbed extragalactic X-ray source, like GRB~190114C. Once identified, the trailing charge events were removed from the event list, resulting in clean WT spectra that are usable below 0.7 keV. 
The XRT spectral extraction then proceeded using standard \Swift analysis software included in  HEASOFT software (version 6.25). Grade 0 events were selected to help mitigate pile-up and appropriately sized annular extraction regions were used, when necessary, to exclude pile-up from the core of the WT point spread function (PSF) profile when the source count rate was greater than $\sim 100$ cts s$^{-1}$. PSF and exposure-corrected ancillary response files were created to ensure correct recovery of the source flux during spectral fitting.

We tested three models in the joint spectral fits, a PL, a broken power law (BKNPL), and a smoothly broken power law (SBKNPL). Each model was multiplied by two photoelectric absorption models, one for Galactic absorption (``TBabs'') and another for the intrinsic host absorption  (``zTBabs''). For the Galactic photoelectric absorption model, an equivalent hydrogen column density is fixed to 7.54 $\times$ 10$^{19}$ atoms cm$^{-2}$ \citep{2013MNRAS.431..394W}. We let the equivalent hydrogen column density for the intrinsic host absorption model be a free parameter in the fit, but fixed the redshift to $z = 0.42$.

\begin{figure}[t!]
\begin{center}
\includegraphics[width=1.0\columnwidth]{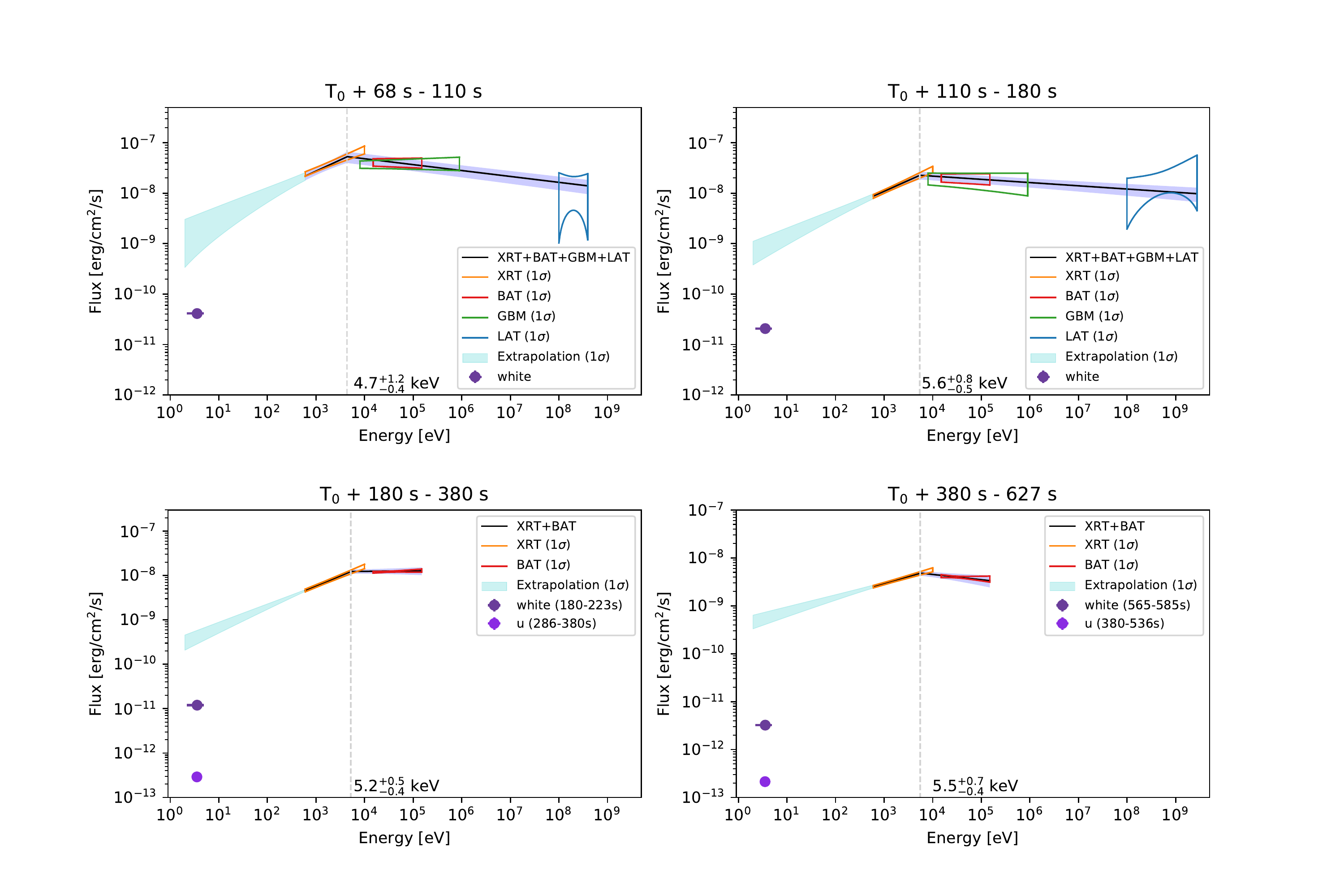}
\caption{Spectral energy distributions from optical to gamma-ray energies for the four time intervals ($T_0$ + 68.27 s to 110 s, $T_0$ + 110 s to 180 s, $T_0$ + 180 s to 380 s and $T_0$ + 380 s to 627 s) described in Section~\ref{sec:analysis:spectral:late}. The solid black lines represent the best-fitting broken power-law function. Each filled region corresponds to the 1-$\sigma$ error contour of the power-law function best-fit to the data from each individual instrument. The cyan regions are an extrapolation from the best-fitting broken power-law function. The dotted line denotes the best-fit break energy $E_{\rm break}$.  The simultaneous UVOT white and $u$ band observations taken during the $T_0$ + 180 s to 380 s and $T_0$ + 380 s to 627 s intervals are also shown, but are not included in the joint spectral fit. Note that the UVOT observations are uncorrected for Galactic or host absorption and as such serve as lower limits to the UV and optical flux.}

\label{late_phase_SED}
\end{center}
\end{figure}

We divided the extended emission phase, $T_{\rm 0}$ + 68.27 s to $T_{\rm 0}$ + 627.14 s, into four time intervals covering 68.27--110 s, 110--180 s, 180--380 s, and 380--627.18 s. The fit results for all four time intervals are listed in Table~\ref{tab:Swift_Fermi}.  For the first two time intervals, we fit the XRT, BAT, GBM, and LAT data simultaneously {by using different fit statistics for each data type: $C_{\rm stat}$ (Poisson data with Poisson background) for the XRT, $\chi^{2}$ for the BAT data, and $PG_{\rm stat}$ for GBM and LAT. These statistics are reported independently for each data set in Table~\ref{tab:Swift_Fermi}.} As shown in Table~\ref{tab:Swift_Fermi} and Figure~\ref{late_phase_SED}, a BKNPL function is statistically preferred over the PL and SBKNPL functions in both time intervals, where Figure~\ref{late_phase_SED} also includes the spectral fitting results using each individual instrument.  When the smoothness parameter $s$ in the SBKNPL model is left free to vary, a sharp break with $s > 10$ is obtained, at which point a SBKNPL resembles a traditional BKNPL model. The low- and high-energy photon indices in the BKNPL model are consistent in both time intervals, yielding $\Gamma_{\rm ph,low}$ $\sim -1.6$ and $\Gamma_{\rm ph,high}$ $\sim -2.1$, respectively, with break energies of 4.22$^{+0.31}_{-0.67}$ keV and 5.11$^{+0.42}_{-0.37}$ keV. We note that the high-energy photon index is consistent with the values in the additional component seen in the prompt phase. This result implies that BAT, GBM, and LAT are observing emission from the same side of the break in the energy spectrum from 10 keV to 100 GeV, which starts to appear during the prompt emission phase in the form of an additional spectral component,  whereas the low-energy channels of the XRT are measuring the energy spectrum below this break.

Since the burst is outside the LAT FoV during the last two time intervals, we limit the joint fit during these intervals to XRT and BAT data. We again simultaneously fit the data to PL and BKNPL models, using again different fit statistics for each data type, $\chi^{2}$ for the BAT data and $C_{\rm stat}$ for the XRT. Again the BKNPL model is statistically preferred over the simpler PL model. For the time interval from $T_{\rm 0}$ + 180 s to 380 s, the low- and high-energy photon indices, as well as the break energy, in the BKNPL model are consistent with those found during the earlier intervals. For the last time interval from $T_{\rm 0}$ + 380 s to 627.14 s, the low-energy photon index is slightly softer than previous intervals, with $\Gamma_{\rm ph,low} = -1.71 \pm 0.05$, and the break energy is almost consistent with previous intervals.

\begin{table*}
\tiny 
\centering 
    \caption{Spectral fitting to \Fermi and \Swift data (1 keV--100 GeV) for various time intervals}
	\begin{tabular}{c c c c c c c c c c c c c}
    \hline\hline
From & To & Model\footnote{Since the XRT data are included, a model is multiplied by the photoelectric absorption models, TBabs with fixed hydrogen column density of 7.54 $\times$ 10$^{19}$ cm$^{-2}$ and zTBabs with fixed redshift of 0.4245.}\footnote{Note that a ``constant'' factor is included to the model, which accounts for the potential of relative calibration uncertainties in the recovered flux (i.e. normalization) between BAT and GBM. The  factor ranges from 0.8 to 1.3, which is acceptable.} &$\Gamma_{\rm ph,low}$ & $\Gamma_{\rm ph,high}$ & $E_{\rm break}$ & \textit{p} & N(H) & $PG_{\rm stat}$ & $C_{\rm stat}$ & $\chi^{2}$ & $dof$ & BIC\\
{[ s ]} & [ s ] & & & & [ keV ] & & [ 10$^{22}$ atoms cm$^{-2}$ ] &  & \\\hline

68.27 & 110 & PL & -2.09$^{+0.01}_{-0.01}$ &  & & & 10.55$^{+0.27}_{-0.26}$  & 504 & 655 & 52 & 1086 & 1239\\
& & BKNPL& -1.55$^{+0.12}_{-0.12}$ & -2.11$^{+0.02}_{-0.02}$ & 4.72$^{+1.20}_{-0.37}$ & & 8.22$^{+0.54}_{-0.52}$ & 502 & 625 & 56 & 1084 & 1225\\
&  & SBKNPL$_{\rm ISM}$\footnote{Smoothness parameter \textit{s} = 1.15 - 0.06\textit{p} \citep{Granot:2002}}& -(\textit{p}+1)/2 & -(\textit{p}+2)/2 & 4.63$^{+3.38}_{-3.28}$ & 2.46$^{+0.08}_{-0.11}$ & 9.91$^{+0.27}_{-0.26}$ & 504 & 642 & 55 & 1085 & 1236\\
&  & SBKNPL$_{\rm wind}$\footnote{Smoothness parameter \textit{s} = 0.80 - 0.03\textit{p} \citep{Granot:2002}}& -(\textit{p}+1)/2 & -(\textit{p}+2)/2 & 7.46$^{+72.89}_{-6.63}$ & 2.54$^{+0.15}_{-0.16}$  & 10.06$^{+0.27}_{-0.26}$ & 504 & 644 & 54 & 1085 & 1238\\\hline 
110 & 180 & PL & -2.00$^{+0.02}_{-0.02}$ &  &  & & 10.42$^{+0.23}_{-0.23}$ & 616 & 671 & 50 & 1087 & 1364\\ 
& & BKNPL & -1.57$^{+0.08}_{-0.08}$ & -2.06$^{+0.02}_{-0.02}$ & 5.60$^{+0.76}_{-0.46}$ & & 8.30$^{+0.40}_{-0.39}$ & 616 & 627 & 51 & 1085 & 1336\\
&  & SBKNPL$_{\rm ISM}$& -(\textit{p}+1)/2 & -(\textit{p}+2)/2 & 2.56$^{+4.20}_{-1.54}$ & 2.24$^{+0.10}_{-0.08}$ & 9.74$^{+0.23}_{-0.23}$ & 621 & 653 & 50 & 1087 & 1358\\
&  & SBKNPL$_{\rm wind}$& -(\textit{p}+1)/2 & -(\textit{p}+2)/2  & 1.69$^{+4.63}_{-0.69}$ & 2.26$^{+0.10}_{-0.10}$ & 9.89$^{+0.23}_{-0.23}$ & 621 & 656 & 50 & 1086 & 1362\\\hline 
 180 & 380 & PL & -1.90$^{+0.01}_{-0.01}$ &  &  & & 9.57$^{+0.17}_{-0.15}$ &  &774 &66 & 810 & 866\\ 
&  & BKNPL & -1.54$^{+0.06}_{-0.06}$ & -1.99$^{+0.05}_{-0.05}$ & 5.18$^{+0.46}_{-0.36}$ & & 7.93$^{+0.29}_{-0.28}$ &  & 727 & 63 & 808 & 830\\
&  & SBKNPL$_{\rm ISM}$&-(\textit{p}+1)/2 & -(\textit{p}+2)/2 & 5.60$^{+0.145}_{-1.512}$ & 2.20$^{+0.10}_{-0.02}$ & 9.07$^{+0.15}_{-0.14}$ &  & 756 & 64 & 809 & 854\\
&  & SBKNPL$_{\rm wind}$&-(\textit{p}+1)/2 & -(\textit{p}+2)/2 & 6.88$^{+0.35}_{-0.44}$ & 2.25$^{+0.16}_{-0.02}$ & 9.21$^{+0.16}_{-0.15}$ &  & 761 & 64 & 809 & 858\\\hline 
380 & 627.14 & PL & -1.86$^{+0.01}_{-0.01}$ &  &  & & 9.09$^{+0.13}_{-0.14}$ &  &700 & 47 & 839 & 775 \\ 
&  & BKNPL& -1.71$^{+0.05}_{-0.05}$ & -2.11$^{+0.08}_{-0.09}$ & 5.52$^{+0.72}_{-0.38}$ & & 8.43$^{+0.24}_{-0.23}$ & & 686 & 42 & 837 & 768\\
&  & SBKNPL$_{\rm ISM}$&-(\textit{p}+1)/2 & -(\textit{p}+2)/2 & 8.67$^{+37.30}_{-6.78}$ & 2.18$^{+0.20}_{-0.16}$ & 8.67$^{+0.20}_{-0.11}$ &  & 694 & 44 & 838 & 772\\
&  & SBKNPL$_{\rm wind}$&-(\textit{p}+1)/2 & -(\textit{p}+2)/2& 9.16$^{+36.96}_{-4.91}$ & 2.20$^{+0.19}_{-0.10}$ & 8.77$^{+0.18}_{-0.09}$ &  & 695 & 45 & 838 & 774\\\hline 

    \end{tabular}
  \begin{tablenotes}
  \item Errors correspond to 1-$\sigma$ confidence region.
  \end{tablenotes}
    \label{tab:Swift_Fermi}
\end{table*}

\subsection{Multiwavelength Afterglow Light Curves} \label{sec:multi-lc}
\begin{figure}[t!]
\begin{center}
\includegraphics[width=0.75\columnwidth]{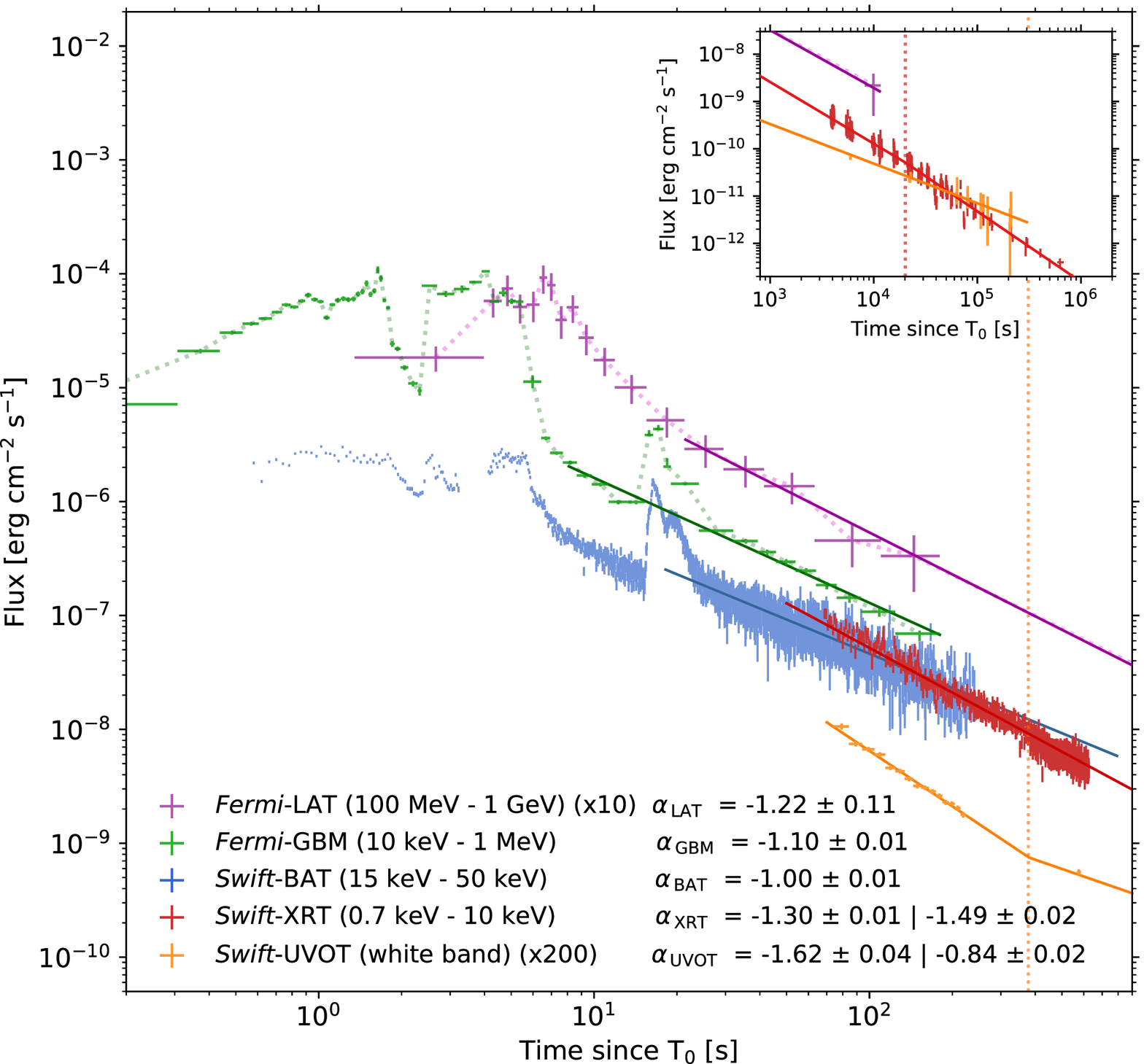}
\caption{Multi-wavelength afterglow light curves for the UVOT (purple), XRT (orange), BAT (red), GBM (green), and LAT (blue) data from GRB 190114C. The flux for the GBM (10 keV -- 1 MeV) and LAT (100 MeV -- 1 GeV) data is calculated from the best-fit model for each time interval in the spectral analysis with each instrument. The BAT, GBM, and LAT emission show a transition after $\sim$ T$_{0}$ + 10 s to an extended emission component decaying smoothly as a power law in time (solid lines). Both the XRT and the UVOT light curves are well described by a broken power law, respectively (solid lines), and their break times are 19.8 $\times$ 10$^3$ s ($\sim$ 5.5 hrs) and 377 s, respectively (dotted lines). The inset shows the light curves of the LAT, XRT and UVOT up to $\sim$ T$_{0}$ + 23 days. }
\label{multi-wavelength_lightcurve}.
\end{center}
\end{figure}

Figure~\ref{multi-wavelength_lightcurve} shows light curves of GRB 190114C for the XRT, BAT, GBM, and LAT data.  The selection for the GBM and LAT data is described in Section~\ref{sec:analysis:spectral:early} and the flux is calculated from the best-fit function for each time interval in the spectral analysis with each individual instrument. The XRT (0.7 keV -- 10 keV), and BAT (15 keV -- 50 keV) light curves are obtained from the UK \Swift Science Data Centre. The UVOT (2 -- 5 eV for the white band) light curve is obtained by {\tt uvotproduct} of HEASoft package. The BAT, GBM, and LAT light curves show an obvious transition from the highly variable prompt emission to a smoothly decaying afterglow component ($\alpha_{\rm BAT} = -1.00 \pm 0.01$, $\alpha_{\rm GBM} = -1.10 \pm 0.01$, and $\alpha_{\rm LAT} = -1.22 \pm 0.11$). At later times, all three light curves decay in time with consistent decay indices, $\alpha$ $\sim -1$, implying that they originate from the same emitting region.

The XRT light curve is well described by a broken power law with temporal indices $\alpha_{\rm XRT}$ of $-1.30 \pm 0.01$ and $-1.49 \pm 0.02$ with the break occurring at approximately $t_{\rm break}$ $\sim$ $T_{0}$ + $\sim$19.8 $\times$ 10$^{3}$ s ($\sim$ 5.5 hrs) (see inset in Figure~\ref{multi-wavelength_lightcurve}). The pre-break decay index of the XRT light curve differs from the indices measured for the BAT, GBM and LAT data.  This difference in decay slopes indicates that the XRT is probing a different portion of the afterglow spectrum, a conclusion that is consistent with the observed spectral breaks in the \Swift and \Fermi joint-fit spectral analysis (Section~\ref{sec:analysis:spectral:late}). 

On the other hand, the UVOT light curve exhibits decay slopes and a temporal break that are distinct from the XRT and BAT data. The temporal break occurs at $\sim$ 400 s, with temporal indices $\alpha_{\rm UVOT}$ before and after the break of $-1.62 \pm 0.04$ and  $-0.84 \pm 0.02$, respectively. These decay indices are steeper than the decay observed in the XRT before the break in the UVOT data and shallower than the XRT decay afterwards. This implies that the UVOT is observing yet another distinct portion of the afterglow spectrum. These observations can be interpreted as the contribution of an optically bright reverse shock that becomes sub-dominant to the forward shock emission at the time of the observed temporal break. In such a scenario, the post-break decay index seen in the UVOT would then reflect a distinct portion of the afterglow spectrum below the X-ray regime.

\section{Discussion} \label{sec:discussion} 

\subsection{Prompt Emission} \label{subsec:prompt}

The prompt emission observed in GRB 190114C resembles the complex relationship between multiple emission components commonly seen in LAT-detected GRBs. The emission observed in the first $\sim2$ s is best characterized as a Band function spectrum with a possible sub-dominant BB component, which combined produce no detectable emission in the LAT energy range. The energy fluxes of the thermal and non-thermal components in the energy band from 10 keV to 1 MeV are $\sim$ 1.1 $\times$ 10$^{-6}$ and $\sim$ 3.9 $\times$ 10$^{-5}$ erg cm$^{-2}$ s$^{-1}$, respectively. We estimate the ratio of the thermal to non-thermal emission during this period to be approximately 3$\%$.

The delay in the onset of the LAT-detected emission is related to the emergence of a hard PL component superimposed on the highly variable Band+BB component seen in the GBM. Furthermore, the PL component is initially attenuated at energies greater than $\sim100$ MeV and we interpret this spectral turnover as due to opacity to electron-positron pair production ($\gamma\gamma \rightarrow e^{+}e^{-}$) within the source. The cutoff energy associated with this turnover is observed to increase with time before disappearing entirely at later times. Similar behavior has been observed in other LAT-detected bursts \citep[e.g.,\ GRB~090926A;][]{Ackermann2011} and has been attributed to the expansion of the emitting region, as the pair production opacity is expected to scale as $\tau_{\gamma\gamma} \propto R^{-1}$ for a fixed mean flux, where $R$ is the distance from the central engine.
 
As has also been noted for other LAT-detected GRBs,  e.g., GRBs 081024B \citep{2010ApJ...712..558A}, 090510 \citep{Ackermann10}, 090902B \citep{2009ApJ...706L.138A}, 090926A \citep{Ackermann+11}, 110731A \citep{Ackermann2013}, and 141207A \citep{2016ApJ...833..139A}, the existence of the extra PL component can be seen as a low-energy excess in the GBM data. This observation disfavors SSC or IC emission from the prompt emission as the origin of the extra PL component, as SSC emission cannot produce a broad power-law spectrum that extends below the synchrotron spectral peak. Instead, we identify this component as the emergence of the early afterglow over which the rest of the prompt emission is superimposed.

\subsection{Afterglow Emission} \label{subsec:afterglow}
The \Swift and \Fermi data reveal that the power-law spectral component observed during the prompt emission transitions to a canonical afterglow component, which fades smoothly as a power law in time. In the standard forward shock model of GRB afterglows \citep{1998ApJ...497L..17S}, specific relationships between the temporal decay and spectral indices, the so-called ``closure relations", can be used to constrain the physical properties of the forward shock as well as the type of environment in which the blast wave is propagating.  

Our broadband fits to the simultaneous XRT, GBM, BAT, and LAT data show evidence for a spectral break in the hard X-ray band (5--10 keV). In the context of the forward shock model, this spectral break could represent either the frequency of the synchrotron emission electrons with a minimum Lorentz factor $\nu_{\rm m}$ or the cooling frequency of the synchrotron emission $\nu_{\rm c}$. Since there are no additional spectral breaks observed up to and through the LAT energy range, if we assume the observed spectral break is either $\nu_{\rm m}$ or $\nu_{\rm c}$, then we naturally hypothesize that $\nu_{\rm c} < \nu_{\rm m}$ or $\nu_{\rm m} < \nu_{\rm c}$, respectively. In the case that the spectral break is $\nu_{\rm m}$, the low-energy and high-energy photon indices are expected to be $\nu^{-1.5}$ for $\nu < \nu_{\rm m}$ and $\nu^{-(p+2)/2} \sim \nu^{-2.1}$ for $\nu > \nu_{\rm m}$, when assuming an electron spectral index of $p \sim$ 2.1.  These values are consistent with the observed photon indices, although the expected temporal index when $\nu < \nu_{\rm m}$ is expected to be $\propto t^{-1/4}$, which is inconsistent with the XRT decay index of $\propto t^{-1.32\pm0.01}$ for either a constant density (ISM) or wind-like (wind) circumstellar environment.  Therefore this scenario in which the break is due to $\nu_{\rm m}$ is disfavored.

In the case that the spectral break is $\nu_{\rm c}$, the low-energy and high-energy photon indices are expected to be $\nu^{-(p+1)/2} \sim \nu^{-1.6}$ for $\nu < \nu_{\rm c}$ and $\nu^{-(p+2)/2} \sim \nu^{-2.1}$ for $\nu > \nu_{\rm c}$, again assuming $p \sim 2.1$, again consistent with the observed values. The expected temporal behavior when $\nu > \nu_{\rm c} $, in both the ISM and wind cases is  $\propto t^{(2-3p)/4} \sim t^{-1.1}$, which is consistent with the temporal decay measured in the BAT, GBM and LAT energy ranges. For $\nu < \nu_{\rm c}$, the expected temporal behavior significantly depends on the density profile of the circumstellar environment.  In the ISM case, the temporal index is expected to be  $\propto t^{3(1-p)/4} \sim t^{-0.8}$, inconsistent with the decay observed in the XRT, whereas for the wind case the expected temporal index is  $\propto t^{(1-3p)/4} \sim t^{-1.3}$, matching the decay seen in X-rays.

If we are indeed observing an afterglow spectrum in which the XRT data are below $\nu_{\rm c}$, then we can follow the formalism established in \cite{2000ApJ...535L..33S} and \cite{2009MNRAS.394.2164V} to estimate an arbitrary circumstellar density profile index $k$, for $n(r) \propto R^{-k}$, to be $k$ =  (12$\beta$ - 8$\alpha$)/(1 + 3$\beta$ - 2$\alpha$) = 1.92 $\pm$ 0.07, which also supports a wind profile ($k$ = 2) scenario.

\begin{figure}[t!]
\begin{center}
\includegraphics[width=0.5\columnwidth]{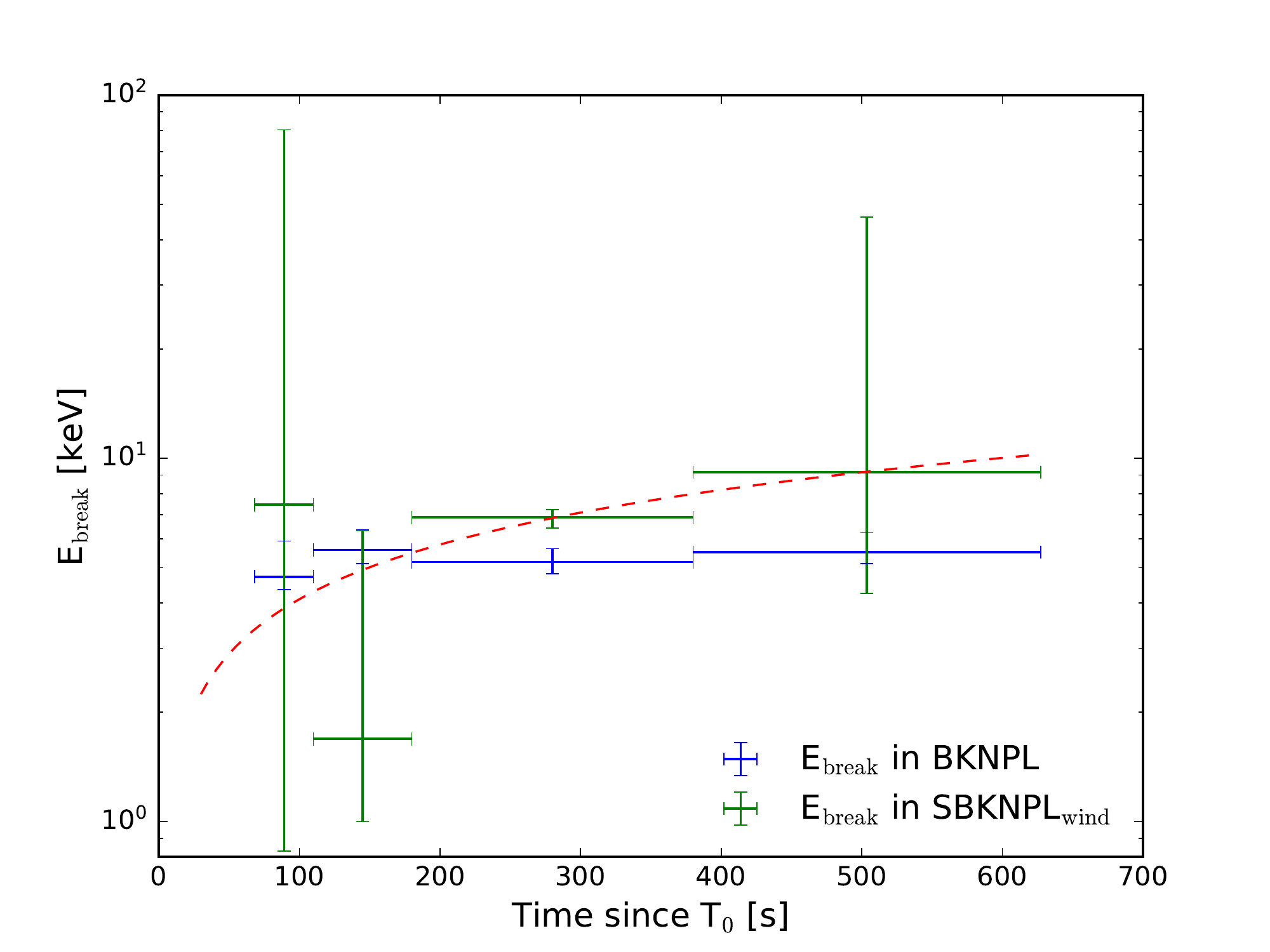}
\caption{Observed spectral break energy versus time. The blue and green points represent the break energy ($E_{\rm break}$) in the BKNPL and SBKNPL$_{\rm wind}$ models in the four time intervals, respectively. The dashed line represents the cooling frequency with time ($\nu_{\rm c} \propto t^{+1/2}$) expected from the afterglow parameters. Despite an initial increase in the break energy between the first two intervals, the break energy is consistent with remaining constant after $T_{\rm 0}$ + $\sim$150 s.}
\label{time_vs_nuc} 
\end{center}
\end{figure}

Figure \ref{time_vs_nuc} shows the observed evolution of $E_{\rm break}$ in the four time intervals we analyzed, along with the expected evolution of the cooling break $\nu_{\rm c} \propto t^{+1/2}$ in a wind-like environment. Despite an initial increase in the break energy between the first two intervals, the break energy is consistent with remaining constant after $T_{\rm 0} > 150$ s. This behavior is similar to that observed for GRB~130427A, in which the broadband modeling preferred a wind-like environment \citep{2014ApJ...781...37P}, but for which $\nu_{\rm c}$ was nonetheless observed to remain constant through the late-time observations \citep{Kouveliotou2013}. \citet{Kouveliotou2013} concluded that GRB~130427A may have occurred in an intermediate environment, possibly produced through a stellar eruption late in the life of the progenitor which altered the circumstellar density profile \citep{2006ApJ...647.1269F}. Nonetheless, a wind-like environment for GRB~190114C matches conclusions drawn by \citet{2011ApJ...732...29C, Ackermann2013, Ajello2018} from a growing number of bursts for a possible preference for LAT-detected bursts to occur in stratified environments, despite the observation that the majority of long GRB afterglows are otherwise consistent with occurring in environments that exhibit uniform density profiles \citep{2011A&A...526A..23S}.

The temporal decay of the UVOT data, although uncorrected for either Galactic or host-galaxy extinction, can provide additional constraints on the location of $\nu_{\rm m}$. The UVOT emission decays as a broken power-law function, starting with $t^{-1.62\pm0.06}$ from 70--400 s, before transitioning to a slower decay of $t^{-0.84\pm0.03}$ for 400--10$^5$ s. The pre-break emission can be interpreted as the contribution from a reverse shock which is expected to exhibit a temporal index of $\propto t^{-(73p+21)/96} \sim t^{-1.82}$, assuming $p$ = 2.1 \citep{2000ApJ...545..807K}, roughly consistent with observations. If the UVOT observed emission after $T_{\rm 0}$ + $\sim$400 s is due to the forward-shock component in which the UVOT data are above $\nu_{\rm m}$ but below $\nu_{\rm c}$, then the temporal decay is expected to be $\propto t^{(1-3p)/4} \sim t^{-1.3}$ for $p$ = 2.1, which is too steep with respect to the observed post-break UVOT decay ($t^{-0.84 \pm 0.03}$). On the other hand, if the UVOT data are below both $\nu_{\rm m}$ and $\nu_{\rm c}$, the temporal decay is expected to be flat, $\propto t^{0}$. Without a clear preference for either of the two scenarios, we conjecture that the UVOT-detected emission may have a different origin or emission site than the X-ray and gamma-ray emission.

\subsection{Energetics}
GRB~190114C was exceptionally bright in the observer frame.  The 1-second peak photon flux measured by GBM is 247 $\pm$ 1 photons s$^{-1}$ cm$^{-2}$, with a total fluence of $(4.433 \pm 0.005) \times 10^{-4}$ erg cm$^{-2},$ both in the 10-1000 keV band. This makes GRB~190114C the fourth brightest in peak flux and the fifth most fluent GRB detected by GBM, placing it in the top $0.3$ percentile of GRBs in the 3rd GBM catalog \citep{GBMBurstCatalog_6Years}. 

\begin{figure}[t!]
\begin{center}
\includegraphics[width=0.5\columnwidth]{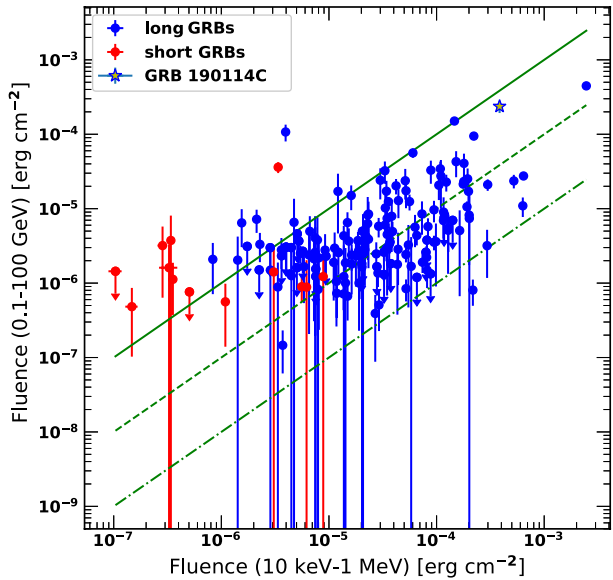}
\caption{Fluence in the energy range of 0.1--100 GeV versus 10 keV--1 MeV  for GRB\,190114C (star) compared with the sample of 186 LAT-detected GRBs from the 2FLGC. Red points are for short GRBs while blue points are for long GRBs.}
\label{fluence} 
\end{center}
\end{figure}

The fluence in the 100 MeV--100 GeV energy band measured by the LAT, including the prompt and extended emission, is (2.4 $\pm$ 0.4) $\times$ $10^{-5}$ erg cm$^{-2}$, which sets GRB\,190114C as the second most fluent GRB detected by the LAT. Figure~\ref{fluence} shows the 10-1000 keV fluence versus the 0.1--100 GeV fluence for GRB\,190114C in comparison with the sample of GRBs detected by the LAT from the 2FLGC. The fluence measured by the LAT is only slightly smaller than that of GRB\,130427A, currently the most fluent GRB detected by the LAT.

At a redshift of $z=0.42$ ($d_{\rm L}=2390$ Mpc), the total isotropic-equivalent  energies $E_{\rm iso}$ released in the rest frame GBM (1 keV--10 MeV), LAT (100 MeV--10 GeV), and combined (1 keV--10 GeV) energy ranges are (2.5 $\pm$ 0.1) $\times10^{53}$ erg, (6.9 $\pm$ 0.7) $\times10^{52}$ erg, and (3.5 $\pm$ 0.1) $\times10^{53}$ erg, respectively. We also estimate a 1-second isotropic equivalent luminosity of $L_{\gamma,{\rm iso}} = (1.07 \pm 0.01 )\times 10^{53} \: {\rm erg} \: {\rm s}^{-1}$ in the 1-10000 keV energy range.

\begin{figure}[t!]
\begin{center}
\includegraphics[width=0.5\columnwidth]{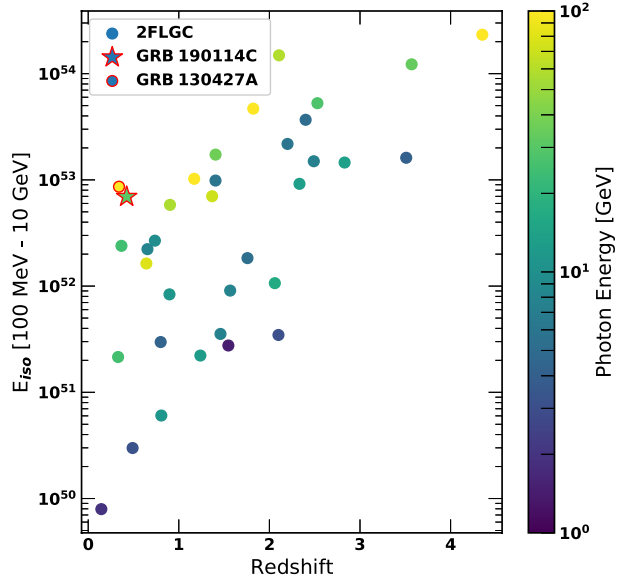}
\caption{Scatter plot of $E_{\rm iso}$ (100 MeV -- 10 GeV) versus redshift for various GRBs including GRB 190114C (star). Colors indicate the energy of the highest-energy photon for each GRB with an association probability $>$90\%. }
\label{energetic} 
\end{center}
\end{figure}

Figure \ref{energetic} shows $E_{\rm iso}$ estimated in the 100 MeV--10 GeV rest frame along with the sample of the 34 LAT-detected GRBs with known redshift in the 2FLGC. 
We note that GRB\,190114C is among the most luminous LAT-detected GRBs below $z < 1$, with an $E_{\rm iso}$ just below GRB\,130427A, which also exhibited the highest-energy photons detected by the LAT from a GRB, including a 95 GeV photon emitted at 128 GeV in the rest frame of the burst.

\subsection{Bulk Lorentz Factor}\label{subsec:bulkLF}

GRBs are intense sources of gamma rays. If the emission originated in a non-relativistic source it would render gamma-ray photons with energies at the $\nu F_{\nu}$-peak energy and above susceptible to $e^\pm$-pair production ($\gamma\gamma\to e^\pm$) due to high optical depths ($\tau_{\gamma\gamma}(\Gamma_{\rm bulk},E)\gg1$) for $\gamma\gamma$-annihilation. This is the so-called `compactness problem' which can be resolved if the emission region is moving ultrarelativistically, with $\Gamma_{\rm bulk}\gtrsim100$, toward the observer \citep{1997ApJ...491..663B, Lithwick-Sari-01,  Granot+08,Hascoet+12}. In this case, the attenuation of flux, which either appears as an exponential cutoff or a smoothly broken power law \citep[][hereafter G08]{Granot+08}, due to $\gamma\gamma$-annihilation occurs at much higher photon energies above the peak of the $\nu F_\nu$ spectrum where $\tau_{\gamma\gamma}(\Gamma_{\rm bulk},E>E_{\rm cut})>1$. Such spectral cutoffs have now been observed in several GRBs, e.g., GRB 090926A \citep{Ackermann+11}; GRBs 100724B and 160509A \citep{Vianello+18}; also see \citet{Tang+15} for additional sources. Under the assumption that these cutoffs indeed result from $\gamma\gamma$-annihilation, they have been used to obtain a direct estimate of the bulk Lorentz factor of the emission region. When no spectral cutoff is observed, the highest energy observed photon is often used to obtain a lower limit on $\Gamma_{\rm bulk}$ instead. In many cases, a simple one-zone estimate of $\tau_{\gamma\gamma}$ was employed, which makes the assumption that both the test photon, with energy $E$, and the annihilating  photon, with energy $\gtrsim \Gamma_{\rm bulk}^2(m_ec^2)^2/E(1+z)^2$, were produced in the same region of the flow \citep[e.g.,][]{Lithwick-Sari-01,Abdo+09}. Such models yield estimates of $\Gamma_{\rm bulk}$ that are typically larger by a factor $\sim 2$ than that obtained from more detailed models of $\tau_{\gamma\gamma}$. The latter either feature two distinct emission regions \citep[a two-zone model;][]{Zou+11} or account for the spatial, directional, and temporal dependence of the interacting photons \citep[G08;][]{Hascoet+12}. Here we use the analytic model of G08 which assumes an expanding ultrarelativistic spherical thin shell and calculates $\tau_{\gamma\gamma}$ along the trajectory of each test photon that reaches the observer. The results of this model have been independently confirmed with numerical simulations \citep{Gill-Granot-18}, which show that it yields an accurate estimate of $\Gamma_{\rm bulk}$ from observations of spectral cutoffs if the emission region remains optically thin to Thomson scattering due to the produced $e^\pm$-pairs. In this case, the initial bulk Lorentz factor of the outflow $\Gamma_{\rm bulk,0}$ is estimated using 
\begin{equation}\label{eq:Gamma-min}
\Gamma_{\rm bulk,0} = 100\left[\frac{396.9}{C_2 (1+z)^{\Gamma_{\rm ph}}}\left(\frac{L_0}{10^{52}\,{\rm erg~s}^{-1}}\right)\left(\frac{5.11\,{\rm GeV}}{E_{\rm cut}}\right)^{1+\Gamma_{\rm ph}}
\left(\frac{-\Gamma_{\rm ph}}{2}\right)^{-5/3}\frac{33.4\,{\rm ms}}{t_v}\right]^{1/(2-2\Gamma_{\rm ph})}~.
\end{equation}
Here $t_v$ is the variability timescale, $\Gamma_{\rm ph}$ is the photon index of the power-law component, 
and $L_0 = 4\pi d_L^2(1+z)^{-\Gamma_{\rm ph}-2}F_0$, where $d_L$ is the luminosity 
distance of the burst, $F_0$ is the (unabsorbed) energy flux ($\nu F_\nu$) obtained at $511\,$keV from the power-law component of the spectrum. The parameter $C_2\approx1$ is constrained from observations of spectral cutoffs in other GRBs \citep{Vianello+18}. The estimate of the bulk Lorentz factor in Eq.(\ref{eq:Gamma-min}) should be compared with $\Gamma_{\rm bulk,max}=(1+z)E_{\rm cut}/m_ec^2$, which corresponds to the maximum bulk  Lorentz factor for a given observed cutoff energy and for which the cutoff energy in the comoving frame is at the self-annihilation threshold, $E_{\rm cut}'= (1+z)E_{\rm cut}/\Gamma_{\rm bulk} = m_ec^2$ 
\citep[however, see, e.g.,][where it was shown that the comoving cutoff energy can be lower than $m_ec^2$ due to Compton scattering by $e^\pm$-pairs]{Gill-Granot-18}. The true bulk Lorentz factor is then the minimum of the two estimates.

In GRB 190114C, the additional power-law component detected by the LAT exhibits a significant spectral cutoff at $E_{\rm cut} \sim 140\,$MeV (where $E_{\rm cut} = E_{\rm pk}/(2+\Gamma_{\rm ph})$) in the time period from $T_{0}$ + 3.8 s to $T_{0}$ + 4.8 s. Using the variability timescale in the GBM band of $t_v \sim$ 6 ms, where we assume that the GBM and LAT emissions are co-spatial, we obtain the bulk Lorentz factor 
$\Gamma_{\rm bulk,0} \sim 210$ from Eq.(\ref{eq:Gamma-min}), which is lower than $\Gamma_{\rm bulk,max}\approx400$ and is therefore adopted as the initial bulk Lorentz factor of the outflow.

\subsection{Forward Shock Parameters} \label{subsec:forward_shock}
The timescale on which the forward shock sweeps up enough material to begin to decelerate and convert its internal energy into observable radiation depends on the density of the material into which it is propagating $A$, the total kinetic energy of the outflow 
($E_{\rm iso}$/$\eta$ $\sim$ 1.8$\times$10$^{54}$ erg, where $E_{\rm iso}$ = 3.5$\times$10$^{53}$ erg $\sim$ 10$^{53.5}$ erg and $\eta$ = 0.2 is the conversion efficiency of total shock energy into the observed gamma-ray emission), and its initial bulk Lorentz factor $\Gamma_{\rm bulk, 0}$. 
Here, in a wind environment, we define a timescale $t_\gamma$ on which the accumulated wind mass is 1/$\Gamma_{\rm bulk, 0}$ of the ejecta mass as 
\begin{equation}
t_{\gamma} = \frac{E_{\rm iso} (1+z)}{16 \pi A m_p c^3  \eta \Gamma_{\rm bulk, 0}^4 }\sim 2\,{\rm s}\,\, A_{\star}^{-1}\left( \frac{E_{\rm iso}}{ 10^{53.5}\,{\rm ergs} }\right)\left( \frac{\eta}{0.2} \right)^{-1} \left( \frac{\Gamma_{\rm bulk, 0}}{200} \right)^{-4}~,
\label{eq:t_onset_wind}
\end{equation}
where $A=3\times10^{35}A_\star~{\rm cm}^{-1}$ with a mass-loss rate 10$^{-5}$ $M_{\sun}$ yr$^{-1}$ in the wind velocity of 10$^{3}$ km s$^{-1}$ for $A_{\star}$ = 1.
If the reverse shock is Newtonian, or at least mildly relativistic (i.e.\ the thin-shell limit; \cite{Sari-Piran-95,2003ApJ...595..950Z}), $t_\gamma$ is the deceleration time $t_{\rm dec}$. In the thin-shell case,
to obtain the observed temporal onset at $T_{\rm 0}$ + $\sim$10 s, $A_\star$ = 0.2 is needed. 
If the reverse shock is relativistic (thick-shell limit), one has $t_{\rm dec} \sim t_{\rm GRB} > t_{\gamma}$ ($t_{\rm GRB}$ is the burst duration), which approximately gives $A_\star >$ 0.2.

Having constrained the location of the synchrotron break energies and the likely environment into which the blast wave is propagating, 
we can invert the equations governing the energies of these breaks to estimate the physical properties of the forward shock.  These 
include the microphysical parameters describing the partition of energy within the shock, the total energy of the shock $E_{\rm K}$ (= $E_{\rm iso}$/$\eta$), and the 
circumstellar density normalization $A_{\star}$. The equations governing 
the location of $\nu_{\rm m}$, $\nu_{\rm c}$, and the flux at which the cooling break occurs $F_\nu(\nu_{\rm c})$ in the case of only synchrotron radiation can be expressed as \citep{2002ApJ...568..820G}:
\begin{equation}
\nu_{\rm c} = 9.1\times10^{11}\, \epsilon_{\rm B}^{-3/2}  \left( \frac{A_{\star}}{0.2}\right)^{-2} \left( \frac{t}{90 \:{\rm s}}\right)^{1/2} \left(\frac{E_{\rm iso}}{10^{53.5} \:{\rm ergs}}\right)^{1/2}\left(\frac{\eta}{0.2 }\right)^{-1/2}~{\rm Hz}
\label{eq:nu_c}
\end{equation}
\begin{equation}
F_\nu(\nu_{\rm c}) = 4.2\times10^{8} \, \epsilon_{\rm e}^{p-1} \epsilon_{\rm B}^{p-1/2}  \left( \frac{A_{\star}}{0.2}\right)^{p}\left( \frac{t}{90 \:{\rm s}}\right)^{1/2-p} \left(\frac{E_{\rm iso}}{10^{53.5} \:{\rm ergs}}\right)^{1/2}\left(\frac{\eta}{0.2 }\right)^{-1/2} ~{\rm mJy}
\label{eq:f_nu_c}
\end{equation}
\begin{equation}
\nu_{\rm m} = 2.1\times10^{19} \, \epsilon_{\rm e}^{2} \epsilon_{\rm B}^{1/2} \left( \frac{t}{90 \:{\rm s}}\right)^{-3/2} \left(\frac{E_{\rm iso}}{10^{53.5} \:{\rm ergs}}\right)^{1/2}\left(\frac{\eta}{0.2 }\right)^{-1/2} ~{\rm Hz}
\label{eq:nu_m}
\end{equation}\textbf{
}
Combining the observed constraints of $\nu_{\rm c} \sim 4$ keV or $9.7 \times 10^{17}$ Hz and $F_\nu(\nu_{\rm c}) \sim 5~{\rm mJy}$ at $T_{\rm 0}$ + 90 s, and the estimated $A_{\star}= 0.2$ assuming the thin-shell case, we estimate the fraction of energy in the magnetic fields $\epsilon_B$ to be 9.9$\times$10$^{-5}$, the fraction of energy in the accelerated electrons $\epsilon_e$ to be 4.0$\times$10$^{-2}$, and $\nu_{\rm m}$ to be $\sim 4 \times 10^{14}$ Hz ($\sim$ 2 eV), which approximately corresponds to the white band of the UVOT. Note that these estimates are derived without taking into account the effect of SSC emission.  These parameters allow us to calculate the expected evolution of the synchrotron cooling frequency with time, which is shown in Figure \ref{time_vs_nuc}, roughly matching the temporal evolution of the observed spectral break in the broadband data.  In the thick-shell case with $A_\star >$ 0.2, if we use fiducial values as $A_\star$ = 1--10, we obtain $\epsilon_{\rm e}$ = (4.2--4.5)$\times$10$^{-2}$, $\epsilon_{\rm B}$ = (120--5) $\times$10$^{-7}$, and $\nu_{\rm m}$ = (1.3--0.3) $\times 10^{14}$ Hz, respectively.

\subsection{Maximum synchrotron energy}\label{subsec:maximum_sync_energy}
The analysis of our broadband data has shown that the observed spectral and temporal characteristics of the early afterglow emission from GRB 190114C are in good agreement with predictions from synchrotron radiation due to electrons accelerated in an external shock.
The existence of late-time high-energy photons detected by the LAT, though, poses a direct challenge to this interpretation. The electrons in this scenario are accelerated via the Fermi process, in which they gain energy as they traverse from one side of the shock front to the other. The maximum photon energy that can be produced by such electrons is set by equating the electron energy loss timescale due to synchrotron radiation to the Larmor timescale for an electron to execute a single gyration (i.e., the shortest route an electron can take across the shock front), and is considered to be roughly $\nu_{\rm max,rest}$  = 2$^{3/2}$ 27$m_{\rm e} c^2$/(16$\pi h \alpha_{\rm f}$) $\sim$ 100 MeV in the comoving frame where $h$ and $\alpha_{\rm f}$ are the Planck and the fine-structure constants, respectively, independent of the magnetic field strength \citep{Ackermann2014}. In the observer frame, this limit is boosted by the bulk Lorentz factor and becomes $\Gamma_{\rm bulk}\nu_{\rm max,rest}/(1+z)$.

\begin{figure}[t!]
\begin{center}
\includegraphics[width=1.0\columnwidth]{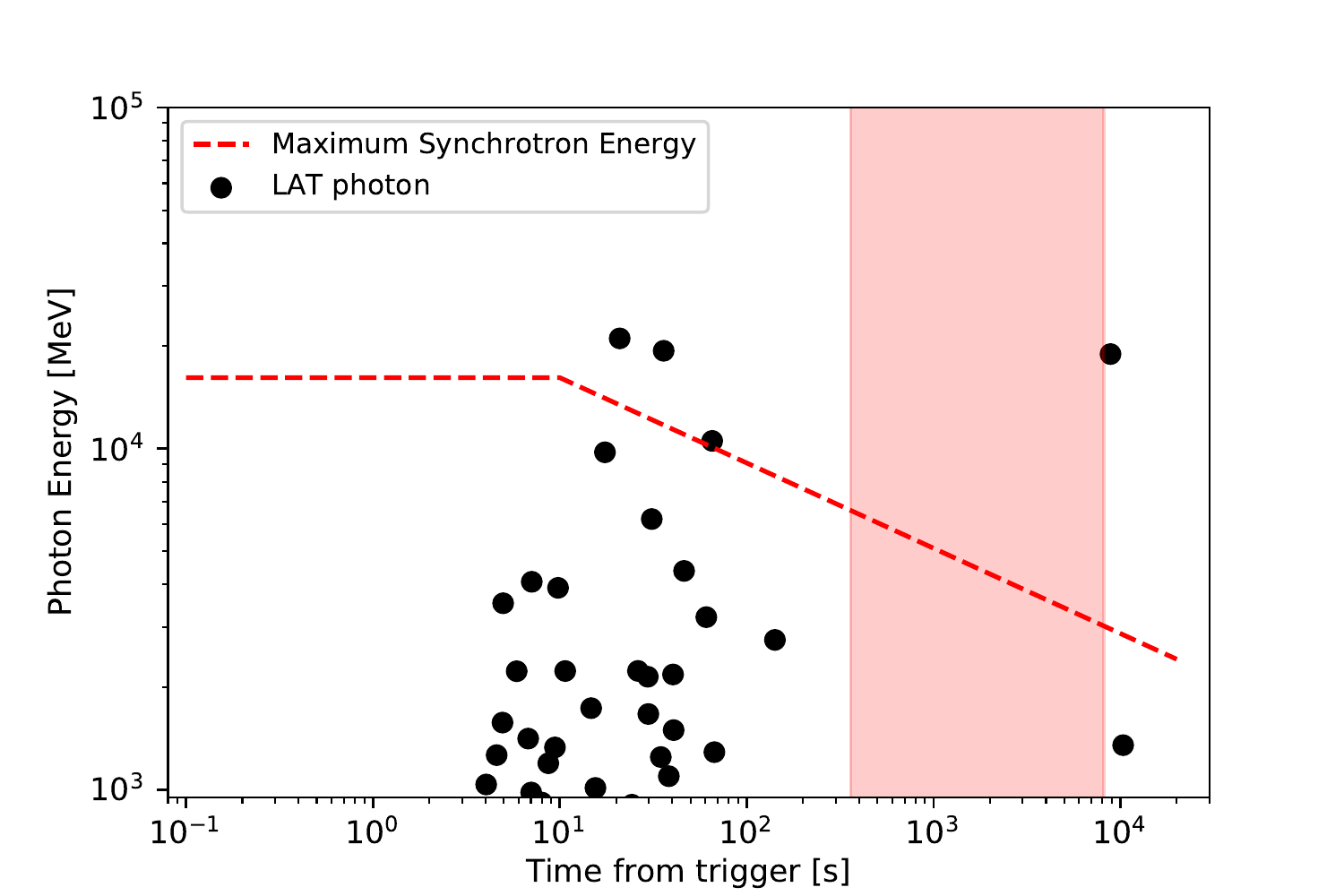}
\caption{Photon energy versus time. Photons with energies $>$ 1 GeV and $>$90\% probability of association with GRB190114C are indicated with black dots. Dashed line represents the maximum synchrotron limit for the adiabatic jet with the wind case. Here we use the estimated bulk Lorentz factor $\Gamma_{\rm bulk}$ = 213, $E_{\rm iso}$ = 3.5 $\times$ 10$^{53}$ erg and the efficiency of total shock energy in converting into the gamma-ray emission $\eta$ = 0.2. The deceleration time for the wind case is calculated with $A_{\star}$ = 0.2.  The red shaded region represents a non-observable period for GRB 190114C due to Earth occultation.}
\label{maximum_syn_photon} 
\end{center}
\end{figure}

We estimated the bulk Lorentz factor at the transition from the coasting to deceleration phases in the previous section. After this transition, the outflow begins to transfer its internal energy to the circumstellar medium and $\Gamma_{\rm bulk}$ of the forward shock decreases with distance from the central engine as $\Gamma_{\rm bulk} \propto R^{-(3-k)/2}$ \citep{1997ApJ...489L..37S}.  As a result, the maximum synchrotron energy decreases with time as the external shock expands.  Using the formalism described in the supplementary material in \cite{Ackermann2014}, we calculate the evolution of $\Gamma_{\rm bulk}(t)$ and use it to estimate the evolution of the maximum synchrotron energy $\nu_{\rm max}(t)$. Figure \ref{maximum_syn_photon} shows the expected maximum synchrotron energy as a function of time along with the observed LAT photons above 1 GeV. Several high-energy photons exceed the expected maximum synchrotron energy at the time of their arrival, including an 18.9 GeV photon arriving approximately 8900 s after $T_{\rm 0}$, almost an order of magnitude higher in energy than our estimate for $\nu_{\rm max}$ at this time.  Given the arrival direction of this photon, we estimate that its association probability with GRB~190114C to be approximately 99.8\%, providing one of the most stringent violations of $\nu_{\rm max}$ observed by the LAT. It is clear that these high-energy detections either necessitate an additional emission mechanism at higher energies, or a revision of the fundamental assumptions used to calculate $\nu_{\rm max}$.

The SSC and IC mechanisms could both produce significant emission above $\nu_{\rm max}$. Synchrotron emission from shock-accelerated electrons should be accompanied by SSC emission, in which the newly created gamma rays gain energy by scattering off energetic electrons before they escape the emitting region. The result is a spectral component that mirrors the primary synchrotron spectrum, but one that is boosted in energy.  In particular, as discussed in Section~\ref{subsec:forward_shock}, for both thin- and thick-shell cases, the observed afterglow parameters indicate a Compton $Y$-parameter of $\epsilon_{\rm e}$/$\epsilon_{\rm B}$ $\sim$ $Y$ $\gg$ 1, in which contributions from the effect of inverse Compton scattering \citep{2000ApJ...543...66P, Sari2001} would be expected. For a bulk Lorentz factor $>$ 100, the peak of the SSC component is expected to be at TeV energies, although as the blast wave decelerates, this peak is expected to evolve into the LAT energy range.  The emergence of such a component should result in a hardening of the LAT spectrum and/or be apparent as deviations in the observed light curve, neither of which has ever been observed in any LAT-detected GRB during their smoothly decaying extended emission. 

One possible solution would require an SSC component to remain sub-dominant to the forward shock synchrotron emission throughout the evolution of the LAT observed emission. Such a scenario could occur when the local energy density of the synchrotron photons is lower than the energy density of the local magnetic field (e.g., \ $Y$ $<$ 1). Furthermore a detailed numerical simulation of the SSC emission considering the evolution of the external-shock emission by \citet{2017ApJ...844...92F} showed that the expected SSC emission could remain weaker than the primary synchrotron emission even if the Compton $Y$-parameter were large. This effect could prevent a significant contribution to the LAT light curve and spectra, while still producing high-energy photons that exceed the maximum synchrotron limit.  

Alternatively, a strong Klein-Nishina (KN) effect could also significantly constrain SSC emission at high energies. This occurs when the energy of the seed photon in the rest-frame of the electrons exceeds $m_{\rm e}c^2$, i.e
$\gamma_{\rm e} E^\prime_{\rm seed} > m_{\rm e}c^2$, where  $\gamma_{\rm e}$ and $E^\prime_{\rm seed}$ are the electron Lorentz factor and the energy of the seed photon in the comoving frame, respectively,
beyond which SSC emission becomes increasingly inefficient. This results in the suppression of high-energy photons, yielding a cutoff in the SSC spectrum. We can estimate the energy at which this cutoff should manifest by reconsidering the forward shock parameter discussed in Section~\ref{subsec:forward_shock} and taking into account SSC and KN effects. Following \citet{2002ApJ...568..820G}, both $\nu_{\rm c}$ from Eq. \ref{eq:nu_c} and $F_\nu(\nu_{\rm c}$) from Eq. \ref{eq:f_nu_c} are multiplied by the factors of (1+$Y$)$^{-2}$ and (1+$Y$)$^{p-1}$. If we consider a case with no KN effect, we find that there are no self-consistent solutions for $\epsilon_{\rm e}$ and $\epsilon_{\rm B}$\footnote{When including the effects of SSC, one finds self-consistent solutions for $\epsilon_{\rm e}$ and $\epsilon_{\rm B}$ only when adopting $A_\star$ $\sim$ 10$^{-3}$: $\epsilon_{\rm e}$ = 1.9$\times$10$^{-1}$, $\epsilon_{\rm B}$ = 4.5$\times$10$^{-3}$, $Y$ = 4.9 for $A_\star$ = 1.3$\times$10$^{-3}$. However, such a very low $A_{\star}$ is not likely for this GRB as discussed in Section~\ref{subsec:forward_shock}. }, emphasizing the need to account for the KN effect when considering the effect of SSC emission. If we assume that the observed $\nu_{\rm c}$ is in the KN regime  (e.g.,\ the observed synchrotron spectrum is unaffected by significant IC losses), then $Y \ll$ 1. Such a scenario would require that the Lorentz factor above which electrons are cooled efficiently, $\gamma_c$, to already be above the Lorentz factor $\hat{\gamma_{\rm c}}$ at which photons cannot be efficiently up-scattered by electrons because they are above the KN limit, where $\hat{\gamma_{\rm c}}$ is given by $m_{\rm e}c^2 \Gamma_{\rm bulk} / h\nu_{\rm syn}(\gamma_{\rm c})$ \citep{2009ApJ...703..675N}. We estimate $\Gamma_{\rm bulk}$ to be $\sim$100 at $T_{\rm 0}+90$ s and $h\nu_{\rm syn}(\gamma_{\rm c})$ to be $\sim$4 keV, which yields $\gamma_{\rm c} > 10^4$. When $\gamma_{\rm m}$ $<$ $\gamma_{\rm c}$ and $\hat{\gamma_{\rm c}} < \gamma_{\rm c}$, high-energy SSC photons are not expected to be strongly damped above energies of $>\Gamma_{\rm bulk}\gamma_{\rm c}m_{\rm e}c^2 \sim 0.5$ TeV. Therefore the LAT-detected photons are not expected to be significantly affected by KN suppression, although the VHE spectrum observed by MAGIC could exhibit curvature due to this effect. 

Revisions to fundamental assumptions about collisionless shock physics have also been put forth to explain apparent violations of the maximum synchrotron energy. \cite{2012MNRAS.427L..40K} showed that the upper limit for synchrotron emission could be raised substantially by relaxing the assumption of a uniform magnetic field in the emitting region.  The authors argue that a magnetic field that decays ahead of the shock front could raise $\nu_{\rm max}$ substantially, but only if the magnetic field gradient varied on a length scale smaller than the distance traveled by the most energetic electrons. This solution could result in a value of $\nu_{\rm max}$ that is orders of magnitude above the canonical estimate and help explain many of the LAT-detected bursts with late-time high-energy photons.

Finally, synchrotron emission above our estimated $\nu_{\rm max}$ could still be possible through contributions from a high-energy hadronic component \citep{Razzaque2010}, or if the electrons were accelerated through a process other than shock acceleration, such as magnetic reconnection, which could act on timescales faster than the Fermi process \citep{1994MNRAS.270..480T, 2001A&A...369..694S, 2003ApJ...597..998L, 2007A&A...469....1G, 2010ApJ...725L.234L, 2015SSRv..191..545K}. The latter scenario can occur in an outflow with a random magnetic field, for example through relativistic turbulence, such that magnetic field dissipation and jet acceleration can occur on a time scale much shorter than the diffusion time \citep{2003ApJ...597..998L, 2009MNRAS.395..472K, 2009ApJ...695L..10L,Granot2016}.

\section{Conclusions} \label{sec:conclusions} 

The joint observations of GRB 190114C by \Fermi and \Swift provide a rich data set with which to examine the complex relationship between prompt and afterglow-dominated emission often observed in LAT-detected GRBs. GRB 190114C is among the most luminous GRBs detected by GBM and LAT below $z < 1$, and exceeded only by GRB~130427A in isotropic-equivalent energy above 100 MeV. Our analysis of the prompt emission shows evidence for both thermal (BB) and non-thermal (CPL or Band) spectral components commonly seen in GRB spectra, in addition to the emergence of an additional PL component extending to high energies that explains the delayed onset of the LAT-detected emission. This additional PL component shows strong evidence for spectral attenuation above 40 MeV in the first few seconds of the burst, before transitioning to a harder spectrum that is consistent with the afterglow emission observed by the XRT and BAT at later times.  We attribute the spectral attenuation of this component to opacity to electron-positron pair production and its evolution to the expansion of the emitting region.  We find that the presence of this extra PL component is also evident as a low-energy excess in the GBM data throughout its evolution, disfavoring SSC or external IC emission from the CPL or Band components as the origin of the extra PL component.   

The long-lived afterglow component is clearly identifiable in the GBM light curve as a slowly fading emission component over which the rest of the prompt emission is superimposed. This allows us to constrain the transitions from internal shock to external shock-dominated emission in both the GBM and the LAT.  The subsequent broadband \Fermi and \Swift data allow us to model the temporal and spectral evolution of the afterglow emission, which is in good agreement with predictions from synchrotron emission due to a forward shock propagating into a wind-like circumstellar environment.  We use the onset of the afterglow component to constrain the deceleration radius and initial Lorentz factor of the forward shock in order to estimate the maximum photon energy attainable through the synchrotron process for shock-accelerated electrons.  We find that even in the LAT energy range, there exist high-energy photons that are in tension with the theoretical maximum photon energy that can be achieved through shock-accelerated synchrotron emission. The detection of VHE emission above 300 GeV by MAGIC concurrent with our observations further compounds this issue and challenges our understanding of the origin of the highest energy photons detected from GRBs. The SSC and IC mechanisms could both produce significant emission above $\nu_{\rm max}$, although as was the case with GRB~130427A, a single power law from X-ray to the LAT energy range is capable of adequately fitting the broadband data, and no significant deviations from a simple power-law decay are evident in the late-time LAT light curve.  We conclude that the detection of high-energy photons from GRB~190114C necessitates either an additional emission mechanism in the LAT energy range that is difficult to separate from the synchrotron component, or revisions to the fundamental assumptions used in estimating the maximum photon energy attainable through the synchrotron process.  The detection of VHE emission from GRBs will be crucial for distinguishing between these two possibilities.

\acknowledgments
 
The \textit{Fermi} LAT Collaboration acknowledges generous ongoing support from a number of agencies and institutes that have supported both the development and the operation of the LAT as well as scientific data analysis. These include the National Aeronautics and Space Administration and the Department of Energy in the United States, the Commissariat \`a l'Energie Atomique and the Centre National de la Recherche Scientifique / Institut National de Physique Nucl¥'eaire et de Physique des Particules in France, the Agenzia Spaziale Italiana and the Istituto Nazionale di Fisica Nucleare in Italy, the Ministry of Education, Culture, Sports, Science and Technology (MEXT), High Energy Accelerator Research Organization (KEK) and Japan Aerospace Exploration Agency (JAXA) in Japan, and  the K.~A.~Wallenberg Foundation, the Swedish Research Council and the Swedish National Space Board in Sweden. Additional support for science analysis during the operations phase is gratefully acknowledged from the Istituto Nazionale di Astrofisica in Italy and the Centre National d'\'Etudes Spatiales in France.

The USRA co-authors gratefully acknowledge NASA funding through contract NNM13AA43C. The UAH co-authors gratefully acknowledge NASA funding from co-operative agreement NNM11AA01A and that this work was made possible in part by a grant of high performance computing resources and technical support from the Alabama Supercomputer Authority. E. B. is supported by an appointment to the NASA Postdoctoral Program at the Goddard Space Flight Center, and C. M. is supported by an appointment to the NASA Postdoctoral Program at the Marshall Space Flight Center, administered by Universities Space Research Association under contract with NASA. C. M. H. and C. A. W.-H. gratefully acknowledge NASA funding through the Fermi GBM project. Support for the German contribution to GBM was provided by the Bundesministerium f¨ur Bildung und Forschung (BMBF) via the Deutsches Zentrum f¨ur Luft und Raumfahrt (DLR) under contract number 50 QV 0301.

This work was performed in part under DOE Contract DE-AC02-76SF00515 and support by JSPS KAKENHI Grant Number JP17H06362, the JSPS Leading Initiative for Excellent Young Researchers program and Sakigake 2018 Project of Kanazawa University (M.A.).

This work made use of data supplied by the UK Swift Science Data Centre at the University of Leicester.

\bibliography{bibliography.bib}

\end{document}